\documentclass[conference,compsoc]{IEEEtran}
\ifCLASSOPTIONcompsoc
  \usepackage[nocompress]{cite}
\else
  \usepackage{cite}
\fi
\ifCLASSINFOpdf
\else
\fi

\usepackage{caption}

\usepackage[linesnumbered,ruled]{algorithm2e}
\usepackage{url}
\usepackage{balance}
\usepackage{enumitem}
\usepackage{tcolorbox}
\usepackage{hyperref}
\usepackage{diagbox}
\usepackage{xspace}
\usepackage{amsmath,amssymb,amsthm}
\usepackage{booktabs}
\usepackage{multirow}
\usepackage{makecell}
\usepackage{float}
\usepackage{marvosym}

\usepackage{colortbl}  
\usepackage{xcolor, eucal}
\usepackage{array}  
\definecolor{gray0}{gray}{0.9}

\newtheorem{theorem}{Theorem}[section]
\newtheorem{definition}{Definition}[section]

\begin{document}
\pagestyle{empty}

\def \toolname{$\mathtt{DP\mbox{-}FETA}$\xspace}
\def \toolnameNone{$\mathtt{DP\mbox{-}FETA_{\uppercase\expandafter{\romannumeral2}}}$\xspace}
\def \toolnameMean{$\mathtt{DP\mbox{-}FETA}_{e}$\xspace}
\def \toolnameMod{$\mathtt{DP\mbox{-}FETA}_{o}$\xspace}
\def \toolnameMeanRaw{$\mathtt{DP\mbox{-}FETA}_{e\mbox{-}r}$\xspace}
\def \toolnameModRaw{$\mathtt{DP\mbox{-}FETA}_{o\mbox{-}r}$\xspace}
\def \toolnamePublic{$\mathtt{DP\mbox{-}FETA}_{pub}$\xspace}

\newcommand\gc[1]{{\color{orange}  #1}}
\newcommand\kc[1]{{\color{black}  #1}}
\newcommand\tw[1]{{\color{blue} [tw: #1]}}
%
\title{From Easy to Hard: Building a Shortcut for Differentially Private Image Synthesis}

\author{\IEEEauthorblockN{Kecen Li\IEEEauthorrefmark{2}\IEEEauthorrefmark{3}\IEEEauthorrefmark{4},
Chen Gong,
Xiaochen Li, 
Yuzhong Zhao\IEEEauthorrefmark{4},
Xinwen Hou\IEEEauthorrefmark{3},
Tianhao Wang
}

\IEEEauthorblockA{University of Virginia, 
\IEEEauthorrefmark{4}University of Chinese Academy of Sciences}

\IEEEauthorblockA{\IEEEauthorrefmark{3}Institute of Automation, Chinese Academy of Sciences}
}

\maketitle

\renewcommand{\thefootnote}{\fnsymbol{footnote}}
\footnotetext[2]{Kecen did this work as a remote intern at UVA.}
\renewcommand{\thefootnote}{\arabic{footnote}}

\begin{abstract}
Differentially private (DP) image synthesis aims to generate synthetic images from a sensitive dataset, \kc{alleviating the privacy leakage concerns of organizations sharing and utilizing synthetic images}. Although previous methods have significantly progressed, especially in training diffusion models on sensitive images with DP Stochastic Gradient Descent (DP-SGD), they still suffer from unsatisfactory performance.

{In this work, inspired by curriculum learning, we propose a two-stage DP image synthesis framework, where diffusion models learn to generate DP synthetic images {\it from easy to hard}. Unlike existing methods that directly use DP-SGD to train diffusion models, we propose an easy stage in the beginning, where diffusion models learn simple features of the sensitive images. To facilitate this easy stage, we propose to use `central images', simply aggregations of random samples of the sensitive dataset. Intuitively, although those central images do not show details, they demonstrate useful characteristics of all images and only incur minimal privacy costs, thus helping early-phase model training.
}
We conduct experiments to present that on the average of four investigated image datasets, the fidelity and utility metrics of our synthetic images are 33.1\% and 2.1\% better than the state-of-the-art method.
The replication package and datasets can be accessed online\footnote{\url{https://github.com/SunnierLee/DP-FETA}}.
\end{abstract}


%
\IEEEpeerreviewmaketitle

\section{Introduction}
\label{sec:intro}

Various previous works proposed that current AI systems face serious security concerns~\cite{gong2024baffle}, and directly conducting statistical analysis on datasets can leak data privacy~\cite{DPText2,DPText3,privsyn}.
An effective approach is to generate synthetic datasets that satisfy \textit{Differential privacy (DP)}~\cite{dp} privacy protections, providing a solution for various statistical tasks~\cite{spas}.
DP offers a theoretical framework to quantify the risk of inferring information about the training dataset from the synthetic dataset, establishing it as a gold standard for privacy preservation~\cite{dp}. 
In recent years, a range of DP data synthesis methods have been proposed, spanning various data types such as tabular data~\cite{privsyn,aim,privmrf}, text~\cite{DPText1,DPText2,DPText3,DPText4}, and images~\cite{privimage,dpdm,api,dp-promise,gong2025dpimagebench}.
These works strive to maintain similarity between the synthetic data and the real dataset while ensuring strict DP guarantees.

This paper focuses on DP image synthesis. One promising approach for DP image synthesis is to train generative models with DP Stochastic Gradient Descent (DP-SGD)~\cite{dpsgd}, which adds Gaussian noise to the training gradients of synthesizers.
Researchers evaluate various generative models such as GANs~\cite{dpsgd-gan1,dpsgd-gan2,dpsgd-gan3,dpsgd-gan4}, diffusion models~\cite{dpdm,dpdm-sota,dpldm,privimage}, and VAEs~\cite{dp2vae,dps-fvae}, among which the diffusion model performs the best.
In particular, Dockhorn et al.~\cite{dpdm} train lightweight diffusion models using a modified DP-SGD, which achieves SOTA performance on standard image synthesis benchmarks. 
However, they still suffer from unsatisfactory performance degradation on some complex image datasets and strong privacy parameters due to the slow convergence of DP-SGD.



Recent studies show that pre-training a generative model with non-sensitive public datasets, which are released on open-source platforms and without privacy concerns, can accelerate the subsequent DP-SGD training and significantly enhance the utility and fidelity of synthetic images~\cite{dpdm-sota,privimage,dpldm,dp-promise}.
\kc{However, previous works reveal that whether the model benefits from pre-training relies on the similarity between the public and sensitive datasets to some extent}~\cite{DPPublicPretraining2,privimage}, which is also verified in our experiments (Section~\ref{subsec:rq2}).
This naturally raises the question: \textit{how to promote DP image synthesis when an appropriate public dataset is not accessible?}

\noindent \textbf{Our Proposal.} 
{Instead of considering how to use an inappropriate public dataset more effectively, we solve this question in another way.}
In Curriculum Learning~\cite{Curriculumlearning},  decomposing complex tasks into multiple steps and learning from easy to hard, are significantly useful in many machine learning tasks~\cite{CurriculumlearningSurvey}. In the DP training of diffusion models, we can apply this idea by breaking down the training process of generating complex images into two stages: (1) the first stage involves training models to learn basic knowledge about the images, such as the general outline, basic color information, and other simple features. We refer to this process as \textit{Warm-up}. After the warm-up, the diffusion models can generate rough and statistically imperfect images. (2) Subsequently, we refine the models to learn the more complex content of the images to generate more realistic images. {We name our proposed framework \toolname, which stands for \textbf{DP} training \textbf{F}rom \textbf{E}asy \textbf{T}o h\textbf{A}rd.}

For the first stage, \toolname obtains some simple features of the sensitive images for diffusion models to learn. To achieve this, we introduce `central images'. Central images are the central tendency measures~\cite{centraltendency} of the sensitive data. Common central tendency measures include mean, mode, median, etc. We find that the central tendency of image data can capture their simple features very well. As shown in Figure~\ref{fig:centerShow}, our two types of central images contain rough outlines of the object and basic color information. The central images are injected with Gaussian noise to ensure DP guarantees.  We warm up the diffusion model by pre-training it on these noisy central images. For the second stage, we fine-tune the diffusion model on the original sensitive images to learn more complex content of the images. To achieve DP, we add Gaussian noise to the model gradient and use the noisy gradient to update the model parameters following standard DP-SGD~\cite{dpsgd}.

\noindent \textbf{Our Evaluations.} We compared our proposed \toolname with existing methods. 
Compared to the state-of-the-art approach using only the sensitive dataset (DPDM~\cite{dpdm}), the fidelity and utility metrics of our synthetic images are 33.1\% and 2.1\% better. \kc{Even when compared to models pre-trained on real public datasets}, our proposed method shows competitive performance, particularly with respect to more `sensitive'\footnote{More `sensitive' means less similar to available public data.} data domains.
{We also find that the central images are effective for warming up diffusion models because they exist in the high probability area of the sensitive dataset. Specifically, we use t-SNE to perform dimensionality reduction on images and find that the distribution of sensitive images is close to that of our queried central images and even closer than that of public images.}

We analyze the impact of the hyper-parameter, specifically the number of queried central images, on the performance. We find that the optimal number of central images is usually much smaller than the number of sensitive images on all our investigated datasets. These results suggest that the warm-up process requires minimal computational resources, presenting the practical applicability in real-world scenarios.

\noindent \textbf{Contributions.} We list our contributions as follows:
\begin{itemize}[leftmargin=*]
    \item {We propose a two-stage DP images synthesis framework, \toolname, where diffusion models learn to generate DP images from easy to hard.}
    \item \kc{Although the warm-up process of \toolname only introduces a minimal amount of computational resource consumption, synthesizers can effectively capture the simple features of sensitive images.}
    \item {Experiments show that \toolname can significantly accelerate the learning of diffusion models and achieves SOTA fidelity and utility metrics on four image datasets without using an additional public dataset.}
\end{itemize}
\vspace{-1mm}
\section{Background}
\label{sec:background}
\vspace{-1mm}


\subsection{Differential Privacy} 
\label{subsec:dp}
\vspace{-1mm}

\noindent \textbf{Definition.}
Differential privacy (DP)~\cite{dp} protects each individual's privacy by requiring any single data in the dataset to have a limited impact on the output. It is defined as follows.
 \begin{definition}[DP~\cite{dp}]
 \label{def:dp} A randomized algorithm $M$ satisfies ($\varepsilon, \delta$)-differential privacy, where $\varepsilon > 0$ and $\delta > 0$, if and only if, for any two adjacent datasets $D$ and $D'$, it holds that,
\begin{equation}
    \Pr[M(D) \in O] \leq e^\varepsilon \Pr[M(D') \in O] + \delta,
    \nonumber
\end{equation}
where $O$ denotes the set of all possible outputs of the algorithm $M$.
\end{definition} The privacy budget $\varepsilon$ is a non-negative parameter that measures the privacy loss in the sensitive data. A smaller $\varepsilon$ indicates better privacy. As usual, we consider ${D,D'}$ are adjacent, denoted $D\simeq D'$, if $D$ can be obtained from $D'$ by adding or removing one element. This paper also uses the above definition to define two neighboring image datasets as previous works~\cite{privimage,datalens}, e.g., one image dataset can be obtained from its neighboring image dataset by adding or removing just one image. 

\noindent\textbf{Sub-sampled Gaussian Mechanism and R\'{e}nyi Differential Privacy.}
This paper uses Sub-sampled Gaussian Mechanism (SGM)~\cite{sgm}, to sanitize central images (introduced in Section~\ref{subsec:stage1}) and use R\'{e}nyi DP (RDP)~\cite{rdp} to track the privacy loss.

\begin{definition}[SGM~\cite{sgm}]
    Let $f:{D_s} \subseteq D \to {\mathbb{R}^d}$ be query function with sensitivity ${\Delta _f} = \max_{D\simeq D'}{\left\| {f\left(D \right) - f\left({D'} \right)} \right\|_2}$.  SGM is parameterized with a sampling rate $q \in \left( {0,1} \right]$ and noise standard deviation $\sigma>0$. The SGM is defined as,
\[
    SG{M_{f,q,\sigma }}\left( D \right) \buildrel \Delta \over = f\left( S \right) + \mathcal{N} \left( {0,{\sigma ^2}\Delta _f^2{\rm I}} \right)
    \nonumber\]
\end{definition}
\noindent where $S =$ \{${x\left| x \in D \right.}$ selected independently with probability $q$\} and $f\left( \varnothing  \right) = 0$.

\begin{definition}[R\'{e}nyi DP~\cite{sgm}]
\label{def:rdp}
    A randomized mechanism $M$ is ($\alpha, \gamma$)-RDP with order $\alpha\in (1, \infty)$, if ${D_\alpha }\left( {M(D)\left\| M(D') \right.} \right) < \gamma$ holds for any adjacent dataset $D$, $D'$, where
    \[{D_\alpha }\left( {Y\left\| N \right.} \right) = \frac{1}{{\alpha  - 1}}\ln {\mathbb{E}_{x\sim N}}{\left[ {\frac{{Y\left( x \right)}}{{N\left( x \right)}}} \right]^\alpha }.\]
\end{definition}
\noindent Then, we obtain the privacy bound of ($\alpha, \gamma$)-RDP by calculating ${D_\alpha }\left( {\left[ \left( {1 - q} \right){p_0} + q{p_1} \right] \left\| {{p_0}} \right. } \right)$. \kc{RDP has a nice linearly composability property: For two mechanisms $M_1$ and $M_2$ satisfying ($\alpha, \gamma_1$)-RDP and ($\alpha, \gamma_2$)-RDP, respectively, the composition ($M_1$, $M_2$) satisfies ($\alpha, \gamma_1+\gamma_2$)-RDP.} RDP can quantify the privacy loss of SGM accurately:
\begin{theorem}[RDP for SGM~\cite{sgm}]
\label{the:rdp4sgm} Let $p_0$ and $p_1$ denote the PDF of $\mathcal{N}(0,\sigma^2 \Delta_f^2)$  and $\mathcal{N}(1,\sigma^2 \Delta_f^2)$ respectively. A $SG{M_{M,q,\sigma }}\left( D \right)$ satisfies ($\alpha, \gamma$)-RDP for any $\gamma$ such that,
    \begin{equation}
    \label{eq:rdp_gamma}
    \gamma  \ge {D_\alpha }\left( {\left[ \left( {1 - q} \right){p_0} + q{p_1} \right] \left\| {{p_0}} \right.} \right).
    \end{equation}
\end{theorem}
\noindent RDP privacy cost $(\alpha,\gamma)$ can be converted to the $(\varepsilon,\delta)$-DP privacy cost as follows.

\begin{theorem}[From $(\alpha,\gamma)$-RDP to $(\epsilon,\delta)$-DP~\cite{rdp}]
\label{the:rdp2dp}
     If $M$ is an ($\alpha, \gamma$)-RDP
mechanism, it also satisfies ($\epsilon, \delta$)-DP, for any $0 < \delta < 1$, where $\epsilon=\gamma + \frac{\log 1/\delta}{\alpha-1}$.
\end{theorem}
Therefore, we can try different $(\alpha,\gamma)$ satisfying Theorem~\ref{the:rdp4sgm} to obtain the smallest $\epsilon$ according to Theorem~\ref{the:rdp2dp} for a tight privacy bound.

\subsection{DP Image Synthesis}
\label{subsec:dp_image}

To generate new images using an available image dataset, the commonly used approach is to query useful information from the training images to estimate the distribution of image data, and then sample new images from the estimated data distribution. For DP image synthesis, where training images are sensitive, the query results used to estimate the data distribution must be injected with suitable noise to satisfy DP. Although previous works have proposed to query the distribution feature~\cite{dp-merf,dp-ntk,pearl,dp-kernel}, they fail to achieve great synthesis performance on complex image datasets. 


Given the success of modern deep generative models, a more promising approach leverages deep generative models to generate DP images. To train a generative model, we optimize the model parameters $\theta$ to minimize a defined objective function $\mathcal{L}$ on a training dataset as,
\begin{align*}
\theta \gets \theta - \eta \left( \frac{1}{|b|} \sum_{i \in b} \mathcal{L}(\theta, x_i)\right),
\end{align*}
where $\eta$ is the learning rate, and $\nabla {\mathcal{L}}(\theta, x_i)$ is the gradient of the loss function ${\mathcal{L}}$ with respect to the model parameters $\theta$ for the data point $x_i$ in a randomly sampled batch $b$ with the sample ratio $q$. Therefore, we can add noise to the gradient of generative models to satisfy DP. A widely adopted method is Differentially Private Stochastic Gradient Descent (DP-SGD)~\cite{dpsgd}, which modifies the parameters update as follows,
{
\begin{align*}
\theta \gets \theta - \eta \left( \frac{1}{|b|} \sum_{i \in b} \text{Clip}\left(\nabla {\mathcal{L}}(\theta, x_i), C\right) + \frac{C}{|b|} \mathcal{N}(0, \sigma^2 {\rm I}) \right),
\end{align*}
}where $\text{Clip}(\nabla {\mathcal{L}}, C) \gets \min\left\{1,\frac{C}{||\nabla {\mathcal{L}}||_2}\right\}\nabla \mathcal{L}$, ${\mathcal{L}}$ refers to a function that clips the gradient vector $ \nabla \mathcal{L} $ such that its $\ell_2$ norm under the constraint of $C$, and \( \mathcal{N}(0, \sigma^2 {\rm I}) \) is the Gaussian noise with the variance $\sigma$. DP-SGD ensures the generative model does not overly learn some specific data points and does not focus on unusual details that might jeopardize privacy.


\subsection{Diffusion Models}
\label{subsec:dm}

Diffusion models~\cite{ddpm,ddpm_song,iddpm} are a class of likelihood-based generative models that learn to reverse a process that gradually degrades the
training data structure. Thus, diffusion models consist of two phases.

\noindent \textbf{Forward Process.} Given an uncorrupted training sample $x_0 \sim p\left(x_0\right)$, diffusion models corrupt $x_0$ by adding Gaussian noise, and output the noised version $\{{x_1}, \ldots ,{x_T}\}$. This process can be obtained according to the following Markov process,
\begin{equation}
\label{DiffusionProcess}
    p\left( {{x_t}\left| {{x_{t - 1}}} \right.} \right) = \mathcal{N}\left( {{x_t};\sqrt {1 - {\beta _t}} {x_{t - 1}},{\beta _t}{\rm I}} \right), \forall t \in \{1, \ldots ,T\},
    \nonumber
\end{equation}
where $T$ is the number of noising steps and ${\beta _t} \in [0,1)$ regulates the magnitude of the added noise at each step. $\rm I$ denotes the identity matrix with the same data dimensions. We denote ${{\bar \alpha }_t}: = \prod\nolimits_{s = 1}^t {\left( {1 - {\beta _s}} \right)} $, and an important property is that the distribution of $x_t$ has another closed form~\cite{ddpm},
\begin{equation}
    p\left( {{x_t}\left| {{x_0}} \right.} \right) = \mathcal{N}\left( {{x_t};\sqrt {{{\bar \alpha }_t}} {x_0},\left( {1 - {{\bar \alpha }_t}} \right){\rm I}} \right).
    \nonumber
\end{equation}
\noindent With this equation, we can sample any noisy version $x_t$ via just a single step as,
\begin{equation}
\label{eq:xtsample}
    {x_t} = \sqrt {{{\bar \alpha }_t}} {x_0} + e\sqrt {1 - {{\bar \alpha }_t}} ,e\sim \mathcal{N}\left( {0,{\rm I}} \right).
\end{equation}
\noindent We usually design a proper $\{{\beta_1}, \ldots ,{\beta_T}\}$ to have $\bar \alpha_T \approx 0$. Thus, as $t$ increases, the data becomes progressively noisier, gradually resembling Gaussian noise more closely and deviating further from the original data sample with each step. 

\noindent \textbf{Reverse Process.} Since the forward process is a Markov process, if we know the noise $e$ at each step, we can generate new data from $p(x_0)$ through progressively denoising a Gaussian noise $x_T$ from $\mathcal{N}\left( {0,{\rm I}} \right)$. Thus, we can train a neural network $e_{\theta}$ parameterized with $\theta$ to predict the noise. Formally, the objective function is defined by~\cite{dpdm} as,
\begin{equation}
\label{eq:L_DM}
{\mathcal{L}_{DM}} = \mathbb{E}_{t \sim \mathcal{U}{\left( {1,T} \right)}} \mathbb{E}_{{x_0}\sim p\left(x_0 \right)}\mathbb{E}_{e \sim \mathcal{N}\left( {0,{\rm{I}}} \right)}{\left\| {e - {e_\theta }\left( {{x_t},t} \right)} \right\|^2}.
\end{equation}
During the generation, we first sample Gaussian noise $x_T$ from $\mathcal{N}\left( {0,{\rm I}} \right)$. With the predicted noise ${e_\phi }\left( {{x_T},T} \right)$ and Equation~\ref{eq:xtsample}, we can estimate clean data $x_0$. Adding noise to $x_0$ following Equation~\ref{eq:xtsample}, we can obtain less noisy data $x_{T-1}$, which can be used to estimate cleaner data $x_0$. Repeating the above process, we use the clean data $x_0$ estimated from $x_1$ as our final synthetic data.
\section{DP-FETA}
\label{sec:method}

\begin{figure}[!t]
    \centering
    \setlength{\abovecaptionskip}{0pt}
    \includegraphics[width=0.98\linewidth]{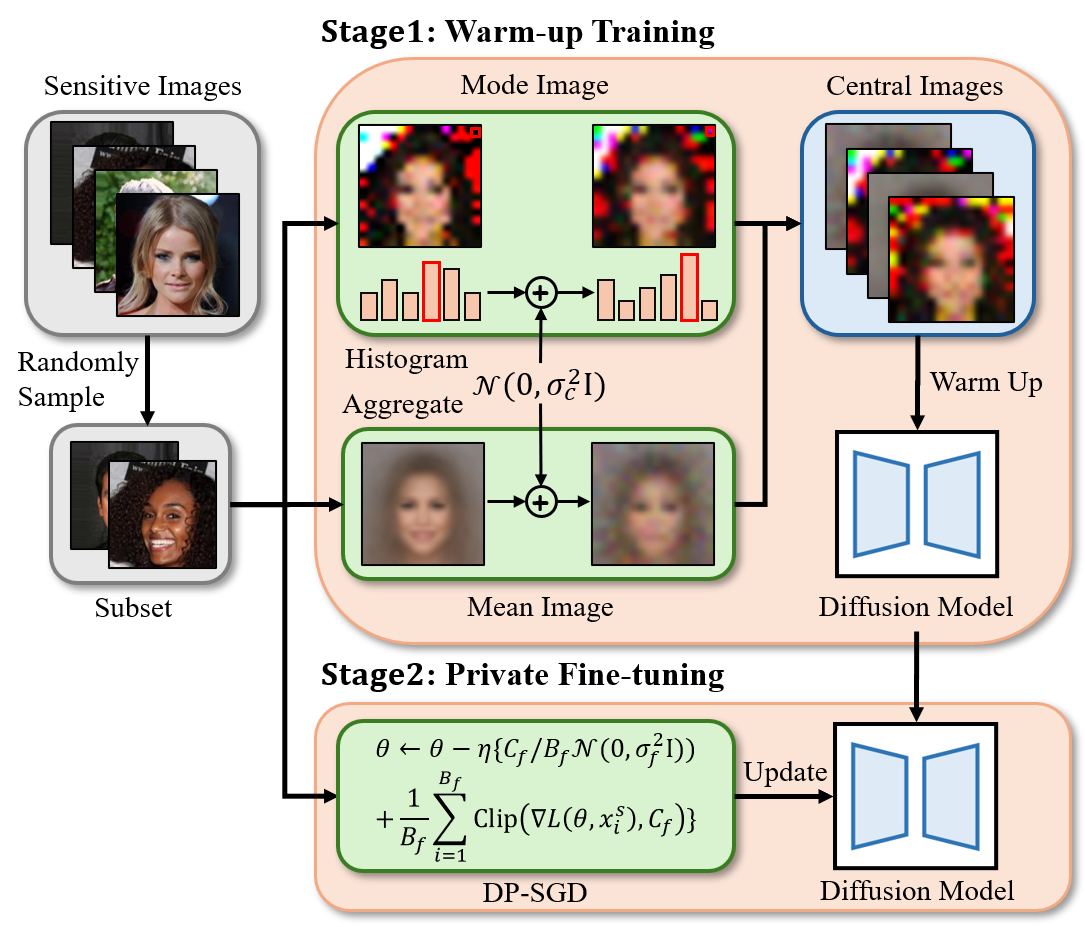}
    \vspace{2mm}
    \caption{The overflow of \toolname. In the first stage, \toolname queries central images from the sensitive images with Gaussian noise injected for DP guarantees. The diffusion model is warmed up on these noisy central images to learn some simple features. In the second stage, the model is fine-tuned with DP-SGD on the original sensitive images to learn more complex features.} 
    \label{fig:framework}
\end{figure}

This section details our proposed \toolname. As shown in Figure~\ref{fig:framework}, \toolname is a two-stage DP image synthesis framework. In the first stage, \toolname queries a central image dataset from the sensitive data with DP guarantees, which is used to warm up the diffusion model to learn some simple features of sensitive images. In the second stage, \toolname fine-tunes the model on the original sensitive images using DP-SGD to generate more realistic images.


\subsection{Stage-\uppercase\expandafter{\romannumeral1}: Warm-up Training}
\label{subsec:stage1}

{In the Stage-\uppercase\expandafter{\romannumeral1}, \toolname aims to construct a central image dataset $D_c = \{x_i^c\}_{i=1}^{N_c}$ consisting of $N_c$ central images $x_i^c$, from the sensitive image dataset $D_s = \{x_i^s\}_{i=1}^{N_s}$ composed of $N_s$ sensitive images, to warm up the diffusion model. We introduce two types of central images, \textit{mean} images and \textit{mode} images, from typical central tendency measures~\cite{centraltendency}. We consider these central images to capture some simple features of the sensitive images, which can be used to warm up diffusion models by learning starting from easy. We first introduce how to construct these two types of central images and then detail how to use these images for warm-up.}

\subsubsection{Mean Images}
\label{subsubsec:mean}

In order to query a mean image, we first sample $B_c$ sensitive images $D_s^{sub}=\{x_i^s\}_{i=1}^{B_c}$ from the sensitive dataset $D_s$ \kc{using Poisson sub-sampling with the sampling probability $q_c$. Similar to DP-SGD~\cite{dpsgd}, $B_c$ is unknown, and we have the expected number $B_c^* = q_c N_s$. We then clip each sensitive image for a controllable bound as follows}:
\begin{equation}
    \label{eq:clip}
    x_i^{s,c} = \min\left\{1,\frac{C_c}{||x_i^s||_2}\right\}\cdot x_i^s,
\end{equation}
\noindent where $C_c$ is a hyper-parameter and the norm of all clipped images is smaller than $C_c$. The mean image is defined as
\begin{equation}
    x^\text{mean} = \frac{1}{B_c^*} \sum_{i=1}^{B_c} {x_i^{s,c}}.
\end{equation}
\noindent We inject suitable Gaussian noise with the noise scale of $\sigma_c$ into the mean image as the following theorem.

\begin{theorem}
\label{the:meanImage}
    The query of mean image $x^\text{mean}$ has global sensitivity $\Delta_\text{mean}=C_c/B_c^*$. For any $\alpha>1$, incorporating noise $\mathcal{N}\left(0,{\sigma_c^2 \Delta_\text{mean}^2} \mathbb{I} \right)$ into the mean image $x^\text{mean}$ makes the query results satisfy $\left(\alpha, \gamma \right)$-RDP, where $\gamma  \ge {D_\alpha }\left( [ \left( {1 - q_c} \right){p_0} + q_c{p_1}] \left\| {{p_0}} \right. \right)$.
\end{theorem}

\noindent We put the proof of Theorem~\ref{the:meanImage} in Appendix~\ref{apsec:MissingProofs}. Therefore, the final mean image $\Tilde{x}^\text{mean}$ is defined as 
\begin{equation}
    \label{eq:meanImage}
    \frac{1}{{B_c^*}} \sum_{i=1}^{B_c} {\min\left\{1,\frac{C_c}{||x_i^s||_2}\right\}\cdot x_i^s} + \mathcal{N}\left(0,{\sigma_c^2 \Delta_\text{mean}^2} \mathbb{I} \right).
\end{equation}
\noindent Repeating the above process $N_c$ times, we can obtain the noisy mean image dataset. Algorithm~\ref{alg:mean} elaborates the process of constructing the noisy mean image dataset.

\begin{algorithm}[!t]
       \caption{Query Mean Image}
      \label{alg:mean}
       \SetKwInOut{Input}{Input}
      \SetKwInOut{Output}{Output}
      \SetKwFunction{meanImageQuery}{meanImageQuery}
      \SetKwProg{Fn}{Function}{:}{}
    \Input{Sensitive dataset $D_s$, number of mean images $N_c$, noise scale $\sigma_c$, size of sample subset ${B_c}$, image norm bound $C_c$.}
    \Output{Noisy mean image set $D_c$}
    \Fn{\meanImageQuery{$D_s$, $N_c$, $\sigma_c$}}{
        Init central dataset $D_c = \varnothing$\;
        \While{$\text{len}\left(D_c\right) < N_c$}{
        Sample subset $\{x_i^s\}_{i=1}^{B_c}$ from $D_s$\\
        Clip images $x_i^{s,c} = \min\left\{1,\frac{C_c}{||x_i^s||_2}\right\}\cdot x_i^s$\\
        \tcp{Aggregation}
         Calculate mean $x^\text{mean} = \frac{1}{B_c^*} \sum_{i=1}^{B_c} {x_i^{s,c}}$\\
         Add noise $\Tilde{x}^\text{mean} = x^\text{mean} + \mathcal{N}\left(0,{\sigma_c^2 C^2 / {B^*_c}^2} \mathbb{I} \right)$\\
         $D_c = D_c \cup \{\Tilde{x}^\text{mean}\}$\\
        }
         \KwRet: $D_c$ 
        }
    \end{algorithm}

\subsubsection{Mode Images}
\label{subsubsec:mode}
Similar to querying mean images, \kc{we first use Poisson sub-sampling with the sampling probability $q_c$ to sample ${B_c}$ sensitive images} from sensitive images $D_s$: $D_s^{sub}=\{x_i^s\}_{i=1}^{B_c}, x_i^s \in \mathbb{R}^{W \times H \times C_x}$, where $W$ and $H$ are the width and height of the image respectively, and $C_x$ is the number of color channels (e.g., its resolution is $W \times H$). However, it is not feasible to directly query the mode image of $D_s^{sub}$ like \toolnameMean for its extremely large global sensitivity. Therefore, we propose to query the histogram of each pixel to obtain the final mode image. For simplicity, we introduce how to query the mode value of one dimension in the image, which can be easily scaled to querying the whole mode image.

Given a set of pixels\footnote{We use one pixel to represent the value in one dimension of the image.} $D_p = \{p_i\}_{i=1}^{B_c}$, which contains ${B_c}$ pixels $p_i \in [0,p_\text{max}]$, we first get its frequency histogram $H_p \in [0,{B_c}]^{\text{bins}}$. $H_p[k]$ represents the number of pixels belonging to $((k-1) \cdot \frac{p_\text{max}}{\text{bins}}, k \cdot \frac{p_\text{max}}{\text{bins}}]$, where $k \in \{1, \ldots, \text{bins}\}$. \kc{The value of `bins' is a hyperparameter, which divides $[0,p_\text{max}]$ into equal parts.} For example, commonly used unsigned 8-bit RGB images have 256-pixel values and $p_\text{max}=255$. To satisfy DP, we inject Gaussian noise with the noise scale of $\sigma_c$ into the frequency histogram using the following theorem.

\begin{theorem}
\label{the:modePixel}
    The query of frequency histogram $H_p$ has global sensitivity $\Delta_\text{p}=1$. For any $\alpha>1$, incorporating noise $\mathcal{N}\left(0,{\sigma_c^2} \mathbb{I} \right)$ into the frequency histogram $H_p$ makes the query results satisfy $\left(\alpha, \gamma \right)$-RDP, where $\gamma  \ge {D_\alpha }\left( \left[ \left( {1 - q_c} \right){p_0} + q_c{p_1} \right] \left\| {{p_0}} \right. \right)$.
\end{theorem}

\begin{figure}[!t]
    \centering
    \setlength{\abovecaptionskip}{0pt}
    \includegraphics[width=1\linewidth]{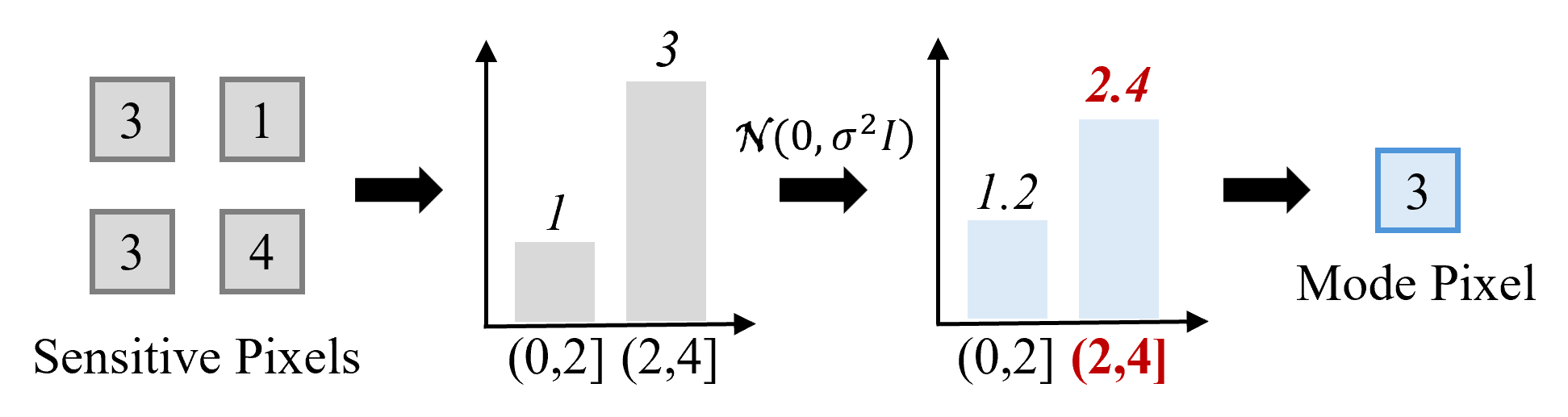}
    \caption{An example of querying the mode pixel from the pixel set $\{1, 3, 3, 4\}$.}
    \label{fig:mode_example}
\end{figure}

\noindent We put the proof of Theorem~\ref{the:modePixel} in Appendix~\ref{apsec:MissingProofs}. Given the noisy frequency histogram $\Tilde{H}_p = H_p + \mathcal{N}\left(0,{\sigma_c^2} \mathbb{I} \right)$, the final mode pixel value is obtained as,
\begin{equation}
    \label{eq:modePixel}
    \Tilde{p}^\text{mode} = \frac{k^* + k^* -1}{2} \cdot \frac{p_\text{max}}{\text{bins}}  = \frac{2k^*-1}{2} \cdot \frac{p_\text{max}}{\text{bins}},
\end{equation}

\noindent where $k^* = \text{argmax}(\Tilde{H}_p)$ is the maximum index in $\Tilde{H}_p$. 
\kc{Consider that the pixel values range from 0 to $p_\text{max}$, divided into $\text{bins}$ equal intervals (or bins). Each bin represents a subrange of pixel values, and $\Tilde{H}_p$ counts the frequency of pixels in each bin, perturbed by noise. The width of each bin is $\frac{p_\text{max}}{\text{bins}}$, meaning that the bin $k$ covers the range $\left[(k-1) \cdot \frac{p_\text{max}}{\text{bins}}, k \cdot \frac{p_\text{max}}{\text{bins}}\right)$. The index $k^*$ identifies the bin with the highest frequency, and we aim to estimate the mode pixel value as a representative point within this bin. The term $\frac{k^* + k^* - 1}{2} = \frac{2k^* - 1}{2}$ computes the midpoint of bin $k^*$ in terms of the bin indices. This midpoint index is then scaled by $\frac{p_\text{max}}{\text{bins}}$, the bin width, to convert it into the actual pixel value scale. Thus, $\Tilde{p}^\text{mode}$ represents the central pixel value of the bin with the highest noisy frequency, approximating the true mode under noise.}

We provide an example of querying the mode pixel in Figure~\ref{fig:mode_example}. 
\kc{We set $p_\text{max}=4$ and $\text{bins}=2$. The pixel $1$ belongs to $(0, 2]$, and $3, 3, 4$ belong to $(2, 4]$. After obtaining the noisy frequency histogram, we have the index of its maximum 2.4,  $k^*=2$. Following Equation~\ref{eq:modePixel}, we have the mode pixel $\Tilde{p}^\text{mode} = \frac{2\times2-1}{2} \cdot \frac{4}{2} = 3$. }To query the whole mode image, we just need to query the frequency histogram of all dimensions $H 
 \in [0,{B_c}]^{W \times H \times C_x \times \text{bins}}$ and sanitize $H$ with the following theorem,

\begin{theorem}
\label{the:modeImage}
    The query of frequency histogram of all dimensions $H$ has global sensitivity $\Delta_\text{mode}=\sqrt{WHC_x}$. For any $\alpha>1$, incorporating noise $\mathcal{N}\left(0,{WHC_x \sigma_c^2} \mathbb{I} \right)$ into the frequency histogram $H$ makes the query results satisfy $\left(\alpha, \gamma \right)$-RDP, where $\gamma  \ge {D_\alpha }\left( {\left[ {\left( {1 - q_c} \right){p_0} + q_c{p_1}\left\| {{p_0}} \right.} \right]} \right)$.
\end{theorem}

We provide the proof of Theorem~\ref{the:modeImage} in Appendix~\ref{apsec:MissingProofs}. After querying the final frequency histogram $\Tilde{H} = H + \mathcal{N}\left(0,{WHC_x \sigma_c^2} \mathbb{I} \right)$, we obtain all the mode pixel values, which compose our mode image. By repeating the query $N_c$ times, we can get $N_c$ noisy mode images. Algorithm~\ref{alg:mode} elaborates on the process of querying the mode image dataset. Since mode images are calculated from the histogram, which needs to discretize the sensitive data, we consider that mode images could better capture the information of simple images. For example, for a black-and-white image (e.g., {\tt MNIST}) where values take only 0 or 255, the mode values belong to $\{0,255\}$, while the mean values might deviate significantly from 0 or 255.

If data labels are available, which usually hold in the conditional generation task, we can partition $D_s$ into multiple disjoint subsets based on labels and query a central dataset for each subset. \kc{Specifically, we group $D_s$ by category, then extract the central images for each subset, representing its key features with minimal privacy cost. The DP guarantee is derived using the parallel composition property~\cite{dp}, which ensures that querying disjoint subsets independently maintains privacy.} All resulting central images are then used for the subsequent warm-up. Figure~\ref{fig:centerShow} presents examples of central images, which capture simple features, such as the general outline of the face.

\begin{algorithm}[!t]
       \caption{Query Mode Image}
      \label{alg:mode}
       \SetKwInOut{Input}{Input}
      \SetKwInOut{Output}{Output}
      \SetKwFunction{modeImageQuery}{modeImageQuery}
      \SetKwProg{Fn}{Function}{:}{}
    \Input{Sensitive dataset $D_s$ where each image $x_i^c \in \mathbb{R}^{W \times H \times C_x}$, number of mean images $N_c$, noise scale $\sigma_c$, size of sample subset ${B_c}$, dimension of histogram `$\text{bins}$'.}
    \Output{Noisy mode image set $D_c$}
    \Fn{\meanImageQuery{$D_s$, $N_c$, $\sigma_c^2$}}{
        \While{$|D_c| < N_c$}{
        \tcp{Query Pixel Histogram}
        Sample subset $\{x_i^s\}_{i=1}^{B_c}$ from $D_s$\\
        \For{$j \gets 1, \ldots, WHC_x$ }{
            \For{$i \gets 1, \ldots, {B_c}$ }{
             Obtain index $k$ from $x_i^s[j]$ \\
             $H[j, k] = H[j, k] + 1$ \\
            }
        }
        Add noise $\Tilde{H} = H + \mathcal{N}\left(0,{WHC_x \sigma_c^2} \mathbb{I} \right)$\\
        \tcp{Obtain Mode Image}
        Init mode image $x^\text{mode} \leftarrow {{\rm O}_{WHC_x \times 1}}$\\
        \For{$j \gets 1, \ldots, W\times H \times C_x$ }{
            $k^* = \text{argmax}(\Tilde{H}_p)$\\
            Mode pixel $\Tilde{p}^\text{mode} = \frac{2k^*-1}{2} \cdot \frac{p_\text{max}}{\text{bins}}$\\
            $x^\text{mode}[j] = \Tilde{p}^\text{mode}$
        }
        
         $D_c = D_c \cup \{\Tilde{x}^\text{mode}\}$\\
        }
         \KwRet: $D_c$ 
        }
\end{algorithm}

\subsubsection{Warm-up}
\label{subsubsec:pre-training}

After querying the central image dataset, we use it to warm up the diffusion models. Specifically, we pre-train models on these central images. However, since the privacy budget is limited, the number of central images we can obtain is small. In our experiments (Section~\ref{subsec:rq2}), we find that diffusion models can easily overfit on these few central images, and the subsequent private fine-tuning can not benefit from the warm-up and even achieves worse performance. Therefore, we consider it important to post-enhance the noisy central image dataset to avoid the overfitting of the warm-up training. In deep learning, there have been many advanced approaches to mitigating overfitting. We adopt data augmentation, which has been commonly used in many computer vision tasks~\cite{ImageAugmentation,ImageAugmentation2}.

To formalize our post-enhancement, we first define an augmentation algorithm bag $\mathcal{B}_a = \{\mathcal{A}_i\}_{i=1}^{N_a}$, which contains $N_a$ different non-deterministic augmentation algorithms, and each algorithm $\mathcal{A}_i$ transforms an input image into a different one. {A naive way is to use each of these algorithms to augment each central image. However, this still produces only a small number of images, since there are very few central images. We consider sequentially augmenting central images. Formally, during the warm-up, given a central image $x^c$, we augment it as follows,
\begin{equation}
\label{eq:aug}
   x_a^c = \mathcal{A}\left(x^c\right) = \mathcal{A}_{a_1} \circ \cdots \circ \mathcal{A}_{a_k} \left(x_i^c\right), a_i \in \{1 \ldots N_a\}.
\end{equation}
This equation means that we randomly sample $k$ augmentation algorithms from $\mathcal{B}$ to sequentially transform the input central image. According to the post-processing mechanism~\cite{dp}, using the noisy central images for warm-up will not introduce any additional privacy cost. Therefore, we can warm up our diffusion models using normal training algorithms for any needed iterations.

\subsection{Stage-\uppercase\expandafter{\romannumeral2}: Private Fine-tuning}
\label{subsec:stage2}

In the Stage-\uppercase\expandafter{\romannumeral2}, we fine-tune the diffusion model on the original sensitive images to learn more complex content of the images. To achieve DP, we add Gaussian noise to the model gradient and use the noisy gradient to update the model parameters following standard DP-SGD~\cite{dpsgd}. Formally, we sample ${B_f}$ sensitive images $D_s^{sub}=\{x_i^s\}_{i=1}^{B_f}$ from the sensitive dataset $D_s$ with sampling probability $q_f$ and the expected number ${B_f^*} = q_f N_s$. The parameters $\theta$ of the diffusion model are updated as follows

{
\footnotesize
\begin{equation}
\label{eq:dpsgd}
    \theta \gets \theta - \eta \left( \frac{1}{{B_f^*}} \sum_{i = 1}^{{B_f}} \text{Clip}\left(\nabla {\mathcal{L}}(\theta, x_i^s), C_f\right) + \frac{C_f}{{B_f^*}} \mathcal{N}(0, \sigma_f^2 {\rm I}) \right),
\end{equation}
}

\noindent where $\mathcal{L}$ is the objective function of diffusion models, and $\eta$ is the learning rate and $\sigma_f^2$ is the variance of Gaussian noise. $\text{Clip}(\nabla {\mathcal{L}}, C_f) = \min\left\{1,\frac{C_f}{||\nabla {\mathcal{L}}||_2}\right\}\nabla \mathcal{L}$ clips the norm of gradient smaller than the hyper-parameter $C_f$.

\begin{table}[!t]
\footnotesize
    \centering
    \caption{The data split and number of categories of four image datasets used in our experiments.}
    \label{tab:datainfo}
    \begin{tabular}{l|ccccc}
    \toprule
    Dataset & Training & Validation & Test & Category \\
    \hline
    {\tt MNIST} & 55,000& 5,000 & 10,000 & 10\\
    {\tt F-MNIST} & 55,000& 5,000 & 10,000 & 10\\
    {\tt CelebA} & 162,770 & 19,867 & 19,962 & 2\\
    {\tt Camelyon} & 302,436 & 34,904  & 85,054 & 2\\
    \bottomrule
\end{tabular}
\end{table}

Algorithm~\ref{alg:dpp} elaborates on the two-stage process of \toolname. We first query a central image dataset $D_c$ following Algorithm~\ref{alg:mean} and~\ref{alg:mode} for mean and mode images, respectively. This dataset $D_c$ is used to warm up the diffusion model with an augmentation algorithm bag. Second, we fine-tune the model on the original sensitive image dataset $D_s$ using standard DP-SGD. We name \toolname that queries mean images and mode images \toolnameMean and \toolnameMod.

\subsection{Privacy Cost}
\label{subsec:PrivacyCost}

In \toolname, two processes consume the privacy budget: (1) querying the central images for warm-up training
 and (2) fine-tuning the warmed-up diffusion model on the sensitive dataset using DP-SGD. According to Theorem~\ref{the:rdp4sgm}, these two processes satisfy $(\alpha,\gamma_w)$-RDP and $(\alpha,\gamma_f)$-RDP, respectively. Specifically, $(\alpha,\gamma_w)$ is determined by the number of central images $N_c$, sample ratio $q_c$ and noise scale $\sigma_c$ according to Theorem~\ref{the:meanImage} and~\ref{the:modeImage}. $(\alpha,\gamma_f)$ is determined by the fine-tuning iteration $t_f$, sample ratio $q_f$ and noise scale $\sigma_f$~\cite{dpsgd}. According to the RDP composition theorem~\cite{rdp}, \toolname satisfies $(\alpha,\gamma_w+\gamma_f)$-RDP.
 \kc{Although more advanced privacy accounting approaches, such as PRV~\cite{prv}, have been proposed, this paper adopts RDP to ensure a fair comparison with existing methods~\cite{dpdm}. We explore how \toolname performs with a better privacy accounting approach. The experimental results in Appendix~\ref{apsubsec:prv} present that \toolname only gains limited improvement. }

To make \toolname satisfy a given $(\varepsilon,\delta)$-DP, we determine privacy parameters following three steps: (1) We set the number of central images $N_c$, sample ratio $q_c$ and noise scale $\sigma_c$ to obtain the RDP cost $(\alpha, \gamma_w)$ of querying central images following Theorem~\ref{the:meanImage} and~\ref{the:modeImage} according to the type of central image. (2) We fix the fine-tuning iterations $t_f$ and sample ratio $q_f$, and then the RDP cost of DP-SGD is a function of noise scale $\sigma_f$ as $(\alpha, \gamma_f(\sigma_f))$. (3) We try different $\sigma_f$ to obtain the corresponding $(\varepsilon,\delta)$-DP following Theorem~\ref{the:rdp2dp}, until meeting the given privacy budget. 

\begin{algorithm}[!t]
       \caption{\toolname Workflow}
      \label{alg:dpp}
       \SetKwInOut{Input}{Input}
      \SetKwInOut{Output}{Output}
      \SetKwFunction{name}{DP-FETA}
      \SetKwFunction{meanQuery}{meanImageQuery}
      \SetKwFunction{modeQuery}{modeImageQuery}
      \SetKwProg{Fn}{Function}{:}{}
    \Input{Diffusion model $e_\theta$ parameterized with $\theta$, sensitive dataset $D_s$, type of central image $t_c$, number of central images $N_c$, noise scale $\sigma_c$.}
    \Output{Trained diffusion model $e_\theta$}
    \Fn{\name{$e_\theta$, $D_s$, $\sigma_c^2$}}{
        \tcp{Stage-\uppercase\expandafter{\romannumeral1}: Warm-up Training}
        \If{$t_c == \text{`mean'}$}{
        $D_c \gets \meanImageQuery(D_s, N_c, \sigma_c^2)$\\
        }
        \If{$t_c == \text{`mode'}$}{
        $D_c \gets \modeImageQuery(D_s, N_c, \sigma_c^2)$\\
        }
        Pre-train the model $e_\theta$ on $D_c$\\
        \tcp{Stage-\uppercase\expandafter{\romannumeral2}: Private Fine-tuning}
        Fine-tune the model $e_\theta$ on $D_s$ using Eq.~\ref{eq:dpsgd}\\
         \KwRet: $e_\theta$ 
        }
    \end{algorithm}



\begin{table*}[!t]
\renewcommand{\arraystretch}{1.1}
\setlength{\tabcolsep}{5.5pt}
\small
    \centering
    \caption{FID and Acc of \toolname and five baselines on {\tt MNIST}, {\tt F-MNIST}, {\tt CelebA} and {\tt Camelyon} with $\varepsilon=1$ and 10. The best performance values in each column are highlighted using the bold font. \toolnameMean and \toolnameMod denote the two versions of \toolname that query the mean images and the mode images as central images, respectively.}
    \label{tab:rq1}
    \resizebox{1.0\textwidth}{!}{
    \begin{tabular}{l|rcrcrcrc|rcrcrcrc}
    \toprule
    \multirow{3}{*}{Method} & \multicolumn{8}{c|}{$\varepsilon=1$} & \multicolumn{8}{c}{$\varepsilon=10$}\\
    \Xcline{2-17}{0.5pt}
    & \multicolumn{2}{c}{{\tt MNIST}} & \multicolumn{2}{c}{{\tt F-MNIST}} & \multicolumn{2}{c}{{\tt CelebA}} & \multicolumn{2}{c|}{{\tt Camelyon}} & \multicolumn{2}{c}{{\tt MNIST}} & \multicolumn{2}{c}{{\tt F-MNIST}} & \multicolumn{2}{c}{{\tt CelebA}} & \multicolumn{2}{c}{{\tt Camelyon}} \\
    \Xcline{2-17}{0.5pt}
     & \centering FID & Acc & FID & Acc & FID & Acc & FID & Acc & FID & Acc & FID & Acc & FID & Acc & FID & Acc\\
    \hline
    DP-MERF~\cite{dp-merf} & 118.3 & 80.5 & 104.2 & 73.1 & 219.4 & 57.6 & 229.3 & 51.8 & 121.4 & 82.0 & 110.4 & 73.2 & 211.1 & 64.0 & 160.0 & 60.0 \\
    G-PATE~\cite{g-pate} & 153.4 & 58.8 & 214.8 & 58.1 & 293.2 & 70.2 & 328.1 & 56.2 & 150.6 & 80.9 & 171.9 & 69.3 & 305.9 & 70.7 & 294.0 & 64.4\\
    DataLens~\cite{datalens} & 186.1 & 71.2 & 195.0 & 64.8 & 297.7 & 70.6 & 343.5 & 50.0 & 173.5 & 80.7 & 167.7 & 70.6 & 320.8 & 72.9 & 381.6 & 50.0\\
    DP-Kernel~\cite{dp-kernel} & 29.5 & 93.4 & 49.5 & 78.8 & 81.8 & 86.2 & 216.5 & 74.3 & 17.7 & 96.4 & 38.1 & 82.0 & 78.5 & 87.4 & 209.9 & 76.4\\
    DPDM~\cite{dpdm} & 23.4 & 93.4 & 37.8 & 73.6 & 77.1 & 84.8 & 57.1 & 81.1 & 5.0 & 97.3 & 18.6 & 84.9 & 24.0 & 92.1 & 45.8 & 81.7\\
    \hline
    \rowcolor{gray0} \toolnameMod & \textbf{8.2} & 96.4 & 28.8 & 80.4 & 42.8 & 83.8 &  53.0 & 82.2 & \textbf{2.8} & \textbf{98.5} & 13.3 & 86.6 & 17.0 & 93.6 & 44.0 & 82.3\\
    \rowcolor{gray0} \toolnameMean & 8.5 & \textbf{96.5} & \textbf{25.6} & \textbf{82.1} & \textbf{41.5} & \textbf{86.5} & \textbf{52.8} & \textbf{82.8} & 2.9 & \textbf{98.5} & \textbf{12.4} & \textbf{87.1} & \textbf{16.1} & \textbf{93.7} & \textbf{42.5} & \textbf{83.1}\\
    \bottomrule
\end{tabular}
}
\end{table*}

\section{Experimental Setup}
\label{sec:exp_setup}

\noindent \textbf{Investigated Datasets.} We perform experiments on four image datasets, {\tt MNIST}~\cite{mnist}, {\tt Fashion-MNIST} ({\tt F-MNIST})~\cite{fmnist}, {\tt CelebA}~\cite{celeba} and {\tt Camelyon}~\cite{camelyon1}. It is noticed that the investigated datasets are prevalently used in previous works to verify the effectiveness of DP image synthesis methods~\cite{dpdm-sota,api,dpdm}.

{\tt MNIST} contains 60,000 handwritten digits in gray
images, from 0 to 9. Similar to {\tt MNIST}, {\tt F-MNIST} comprises 60,000 images of 10 different products. Compared to {\tt MNIST} and {\tt F-MNIST}, {\tt CelebA} and {\tt Camelyon} are more ``sensitive'' image datasets. {\tt CelebA} contains more than 202,599 face images of 10,177 celebrities, each with 40 attributes. Following previous studies, we choose the ``Gender'' to divide {\tt CelebA} into male and female images. {\tt Camelyon} comprises 455,954 histopathological image patches of human tissue, and all images are labeled whether at least one pixel has been identified as a tumor cell. As presented in Table~\ref{tab:datainfo}, all datasets are divided into a training set, a validation set, and a test set.

\noindent \textbf{Baselines.} This paper selects DP-MERF~\cite{dp-merf}, G-PATE~\cite{g-pate}, DataLens~\cite{datalens}, DPDM~\cite{dpdm} and DP-Kernel~\cite{dp-kernel} as baselines. These DP image synthesis methods generate synthetic image datasets without using public data. The implementations are based on their open-source codes. For more details about these methods, please refer to Appendix~\ref{apsubsec:Baselines}.










\noindent \textbf{Evaluation Metrics.} We respectively evaluate the fidelity and utility of the synthetic dataset using two metrics: Fr\'{e}chet Inception Distance (FID) and downstream classification accuracy (Acc), as commonly employed in prior studies~\cite{dpdm,dp-kernel,gong2025dpimagebench}. Please refer to Appendix~\ref{apsubsec:MetricDetails} for more details.





\noindent \textbf{Implementation.} All image generative methods are realized with Python 3.8 on a server with 4 NVIDIA GeForce A100 and 512GB of memory. We replicate all five baselines using their open-source implementation repositories. For the warm-up training of \toolname, we query 50 central images for {\tt MNIST} and {\tt F-MNIST}, and 500 central images for {\tt CelebA} and {\tt Camelyon}. The number of augmentation algorithms $N_a$ is set as 14. Since DPDM~\cite{dpdm} also trains diffusion models with DP-SGD for image synthesis, we use the same settings of private fine-tuning as DPDM for a fair comparison. We recommend readers refer to Appendix~\ref{apsec:id} for more implementation details.

\section{Results Analysis} 
\label{sec:eval_results}

This section evaluates the effectiveness of \toolname by answering three Research Questions (RQs),
\begin{itemize}[leftmargin=*]
    \item \textbf{RQ1.} Does \toolname outperform the five baseline methods across the four investigated image datasets?
    \item \textbf{RQ2.} How does the warm-up training aid in better training diffusion models with DP-SGD?
    \item \textbf{RQ3.} How do the hyper-parameters of \toolname affect the synthesis performance?
\end{itemize}
It is noticed that \toolnameMean and \toolnameMod denote the two versions of \toolname that query the mean images and the mode images as central images.

\subsection{RQ1: Comparison with Existing Methods}
\label{subsec:rq1}

This RQ explores whether \toolname can generate synthetic images with higher fidelity and utility than baselines. We compare our \toolnameMean and \toolnameMod with five baselines on four investigated image datasets as described in Section~\ref{sec:exp_setup}, under the privacy budget $\varepsilon=\{10,1\}$.

Table~\ref{tab:rq1} shows the FID and Acc of \toolname and five baselines. \toolname outperforms all baselines in terms of the FID and Acc of downstream classification tasks using synthetic images on four investigated datasets. When $\varepsilon=10$, both \toolnameMean and \toolnameMod achieve better synthesis performance. Specifically, compared to the SOTA method DPDM~\cite{dpdm}, on average, the FID and Acc of the synthetic images from \toolnameMean are 28.8\% lower and 1.6\% higher respectively, and \toolnameMod obtains 26.4\% lower FID and 1.3\% higher Acc. As $\varepsilon$ decreases to 1, \toolnameMean still remains the SOTA results and obtains 37.4\% lower FID and 3.8\% higher Acc than DPDM on the average. \toolnameMod achieves the best synthesis results except for the Acc of synthetic {\tt CelebA} images, which is only 1.0\% lower than DPDM. Compared to \toolnameMod using mode images for warm-up training, \toolnameMean, using mean images, achieves better synthesis performance. We explore the reasons in Section~\ref{subsec:MeanVSMode}.

\begin{figure*}[!t]
    \centering
    \setlength{\abovecaptionskip}{0pt}
    \includegraphics[width=0.99\linewidth]{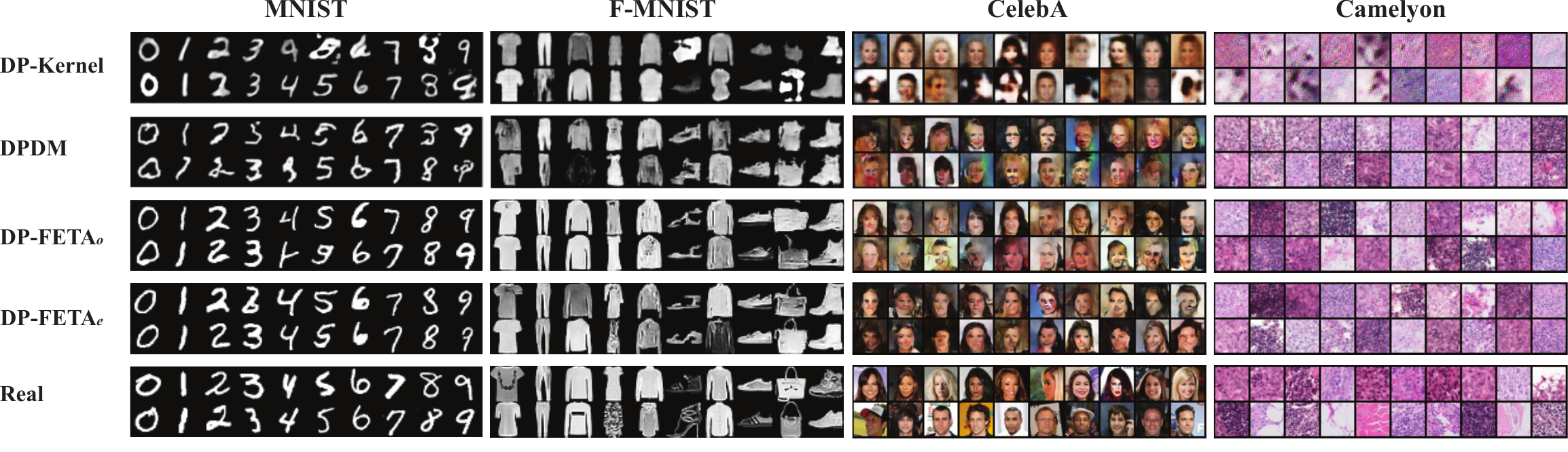}
    \caption{Examples of synthetic images from four different methods, DP-Kernel~\cite{dp-kernel}, DPDM~\cite{dpdm} and our \toolnameMod and \toolnameMean, on four investigated image datasets, {\tt MNIST}, {\tt F-MNIST}, {\tt CelebA} and {\tt Camelyon}, with $\varepsilon=1$. The last row of images are real image samples from each image dataset.}
    \label{fig:rq1}
\end{figure*}

\begin{figure*}[!t]
    \centering
    \setlength{\abovecaptionskip}{0pt}
    \includegraphics[width=1.0\linewidth]{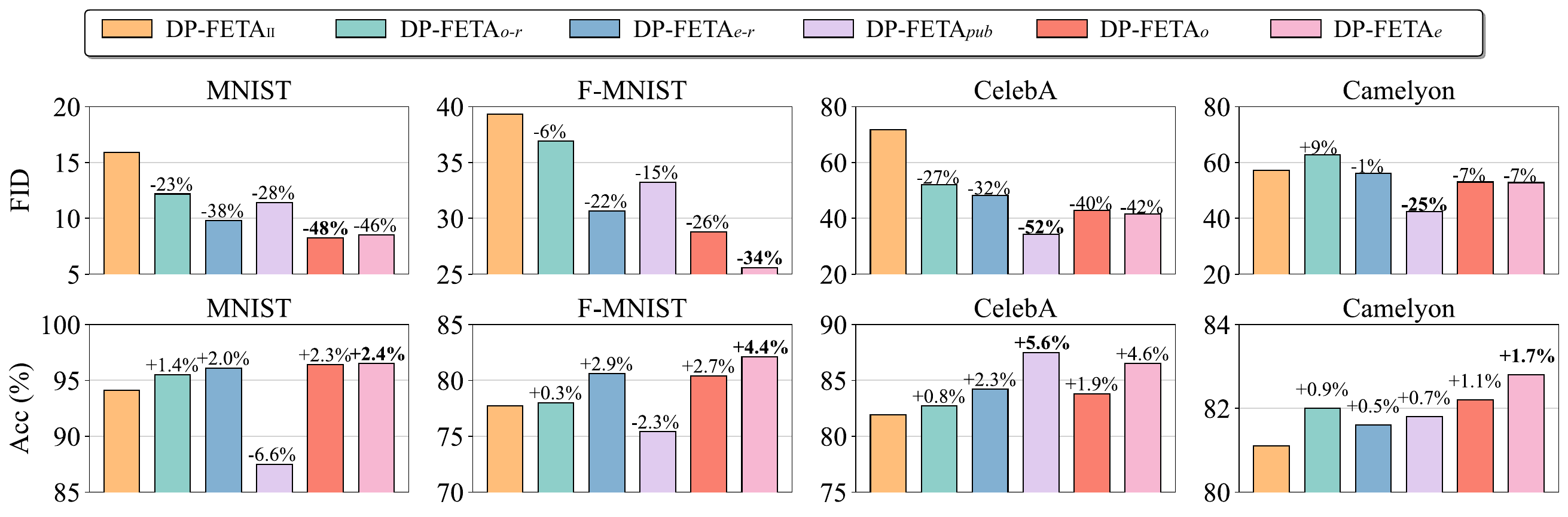}
    \caption{FID (top row) and Acc (bottom row) of \toolnameMean, \toolnameMod and four baselines, which are introduced in  Section~\ref{subsec:rq2}, on {\tt MNIST}, {\tt F-MNIST}, {\tt CelebA} and {\tt Camelyon} with $\varepsilon=1$.}
    \label{fig:rq2}
    \vspace{-2mm}
\end{figure*}

Examples of synthetic images, under $\varepsilon=1$, from various methods are present in Figure~\ref{fig:rq1}. The comparison on $\varepsilon=10$ is put in Figure~\ref{fig:rq1_e10} of the Appendix for space limitation. We only show the top-2 methods, e.g., DP-Kernel and DPDM, in the baselines for space limitation. On four image datasets, both our \toolnameMod and \toolnameMean generate more realistic synthetic images than DP-Kernel and DPDM. Specifically, the generation quality of DP-Kernel is inconsistent, as shown in its synthetic images on {\tt CelebA} and {\tt Camelyon}. The images generated by DPDM are blurry, especially on {\tt F-MNIST}, while the images generated by our \toolnameMod and \toolnameMean have clear object contours. This is because diffusion models with central images for warm-up training can more accurately learn the distribution of training images (e.g., clear contours), while DPDM suffers from slower convergence of DP-SGD and can only learn some relatively coarse image features.

\begin{table}[H]
\normalsize
\setlength{\tabcolsep}{3pt}
    \centering
    \renewcommand\arraystretch{1}
    \begin{tabular}{p{0.97\columnwidth}}
    \Xhline{1.0pt}
         \rowcolor{gray0} \noindent \textbf{Answers to RQ1}: Synthetic images produced by \toolname exhibit greater fidelity and utility compared to all baseline methods with two distinct privacy budgets. On average, the FID and Accuracy (Acc) of the downstream classification task of synthetic images from \toolname is 33.1\% lower and 2.1\% higher than the SOTA method.\\
    \Xhline{1.0pt}
    \end{tabular}
\end{table}

\subsection{RQ2: Effective Warm-up Training}
\label{subsec:rq2}

We explore how our warm-up training improves the DP-SGD training of diffusion models. We introduce four baselines in this experiment as follows:

\begin{itemize}[leftmargin=*]

\item  \textbf{\toolnameNone:} \toolnameNone only involves the second stage of \toolname, and does not query central images to warm up diffusion models, which is the same as DPDM.


\item \textbf{\toolnameMeanRaw:} \toolnameMeanRaw uses the mean images for warm-up, but the images are not post-enhanced by our augmentation algorithm bag.

\item \textbf{\toolnameModRaw:} Similar to \toolnameMeanRaw, \toolnameModRaw warms up diffusion models using the mode images, which are not post-enhanced by the augmentation algorithm bag.

\item \textbf{\toolnamePublic:} \toolnamePublic warms-up diffusion models with a real image datasets, {\tt ImageNet}~\cite{imagenet}, which has been used as the public dataset by previous works~\cite{dpdm-sota,api}.

\end{itemize}

\begin{figure*}[!t]
    \centering
    \setlength{\abovecaptionskip}{0pt}
    \includegraphics[width=0.98\linewidth]{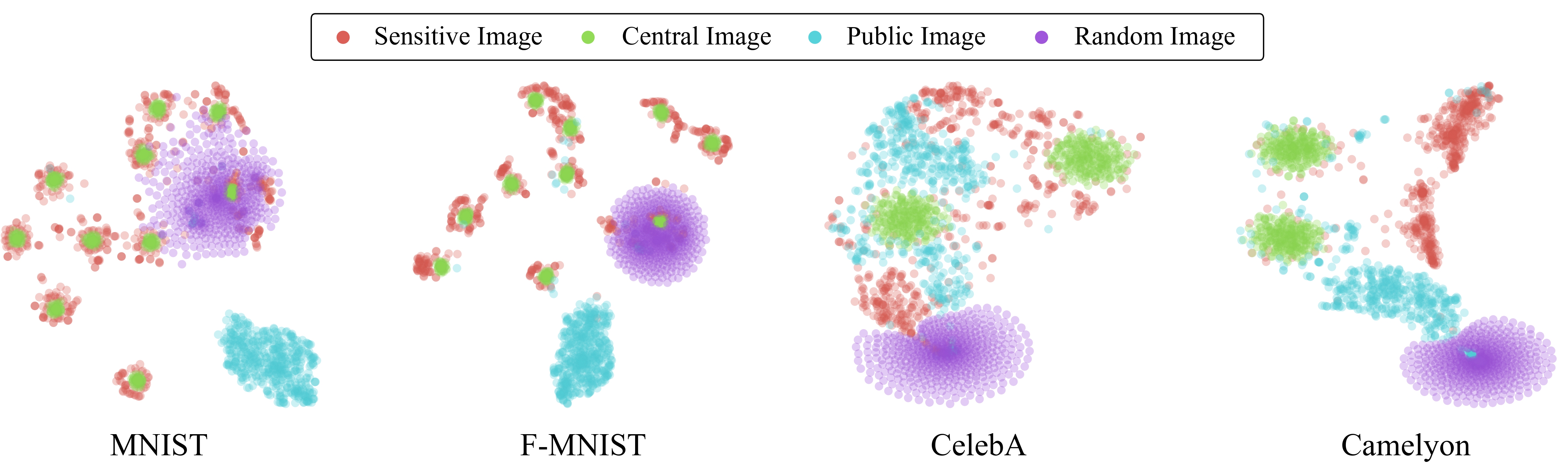}
    \caption{The t-SNE visualizations depict the distribution of the sensitive, queried central, public, and random images.}
    \label{fig:tsne}
    \vspace{-3mm}
\end{figure*}

Figure~\ref{fig:rq2} shows the FID (top row) and Acc (bottom row) of \toolnameMean, \toolnameMod, and four baselines on four investigated datasets with $\varepsilon=1$. Compared to \toolnameNone, which does not involve warm-up training, on average, both \toolnameMean and \toolnameMod achieve better performance with 32.8\% lower FID, 3.3\% higher Acc, and 30.6\% lower FID, 2.0\% higher Acc, respectively. This indicates that it is valuable to allocate a small privacy budget to querying the central images for warm-up training rather than directly allocating the whole privacy budget to the DP-SGD training.

Compared to \toolnameMeanRaw and \toolnameModRaw, which do not post-enhance the queried central images with the augmentation algorithm bag, \toolnameMean and \toolnameMod obtain better FID and Acc. On average, the FID and Acc of synthetic images from \toolnameMean is 
 12.4\% lower and 1.4\% higher than \toolnameMeanRaw, respectively. Consistently, \toolnameMod obtains 21.9\% lower FID and 1.2\% higher Acc than \toolnameModRaw.

Compared to \toolnamePublic, which requires an additional real image dataset for warm-up training, \toolnameMean and \toolnameMod achieve competitive results. On {\tt MNIST} and {\tt F-MNIST}, which contains only gray images, using colorful images from {\tt ImageNet} to warm up diffusion models does not bring great benefits to subsequent fine-tuning, and even has a negative impact on the utility of synthetic images. On average, \toolnamePublic obtains a 21.9\% lower FID, but 4.4\% lower Acc than \toolnameNone. \toolname obtains 22.4\% lower FID and 7.4\% higher Acc than \toolnamePublic on the average of two versions of \toolname. On {\tt CelebA} and {\tt Camelyon}, which are also composed of colorful images, \toolnamePublic achieves competitive performance. On the face image dataset {\tt CelebA}, \toolnamePublic obtains 17.3\% lower FID and 1.0\% higher Acc than \toolnameMean, which may benefit from the face images within {\tt ImageNet}~\cite{privimage}. On {\tt Camelyon}, which consists of human tissue images and differs from {\tt ImageNet} greater~\cite{privimage}, the benefit decreases. Although the FID of synthetic images from \toolnamePublic is still lower than \toolname, the Acc of \toolnamePublic is 0.4\% and 1.0\% lower than \toolnameMod and \toolnameMean, respectively. A natural question is whether we can utilize both central images and public images together to achieve a better synthesis, which will be discussed in Section~\ref{subsec:dis1}.

Additionally, to further validate the effectiveness of our queried central images, we investigate the distribution characteristics of four different types of images as follows:

\begin{itemize}[leftmargin=*]

\item  \textbf{Sensitive Image:} We randomly sample 500 real images from each sensitive image dataset.

\item \textbf{Central Image:} We query 500 mean images from each sensitive image dataset following Equation~\ref{eq:meanImage}.

\item \textbf{Public Image:} We randomly sample 500 real images from the public image dataset {\tt ImageNet}.

\item \textbf{\kc{Random Image}:} We randomly generate 500 Gaussian images to represent the randomly initialized diffusion model. The mean and variance of the Gaussian distribution are set as 0 and 1, respectively.

\end{itemize}

\begin{figure*}[!t]
    \centering
    \setlength{\abovecaptionskip}{0pt}
    \includegraphics[width=1.0\linewidth]{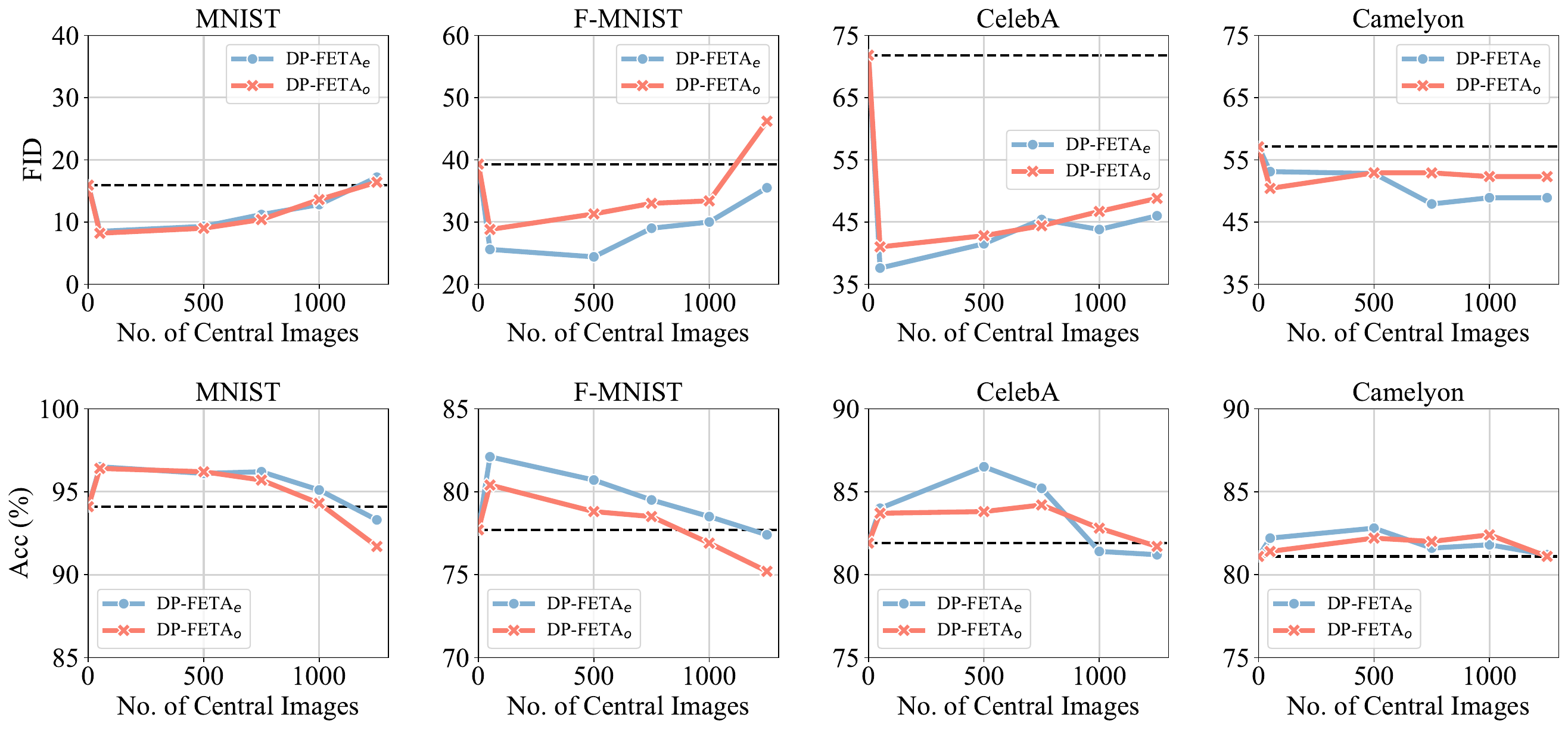}
    \caption{The first and second rows of figures present the effect of different numbers of queried central images on the FID and Acc of \toolname across four investigated datasets with $\varepsilon=1$ respectively. The black dashed line represents the diffusion models without using central images for warm-up training.}
    \label{fig:rq3}
    \vspace{-3mm}
\end{figure*}

We use t-SNE to visualize all images in a two-dimensional space, which is present in Figure~\ref{fig:tsne}.
In this figure, we observe that compared to public images, our central images from {\tt MNIST} and {\tt Fashion-MNIST}—both grayscale datasets—are projected into 10 distinct clusters, closely aligned with the 10 categories of the sensitive images. Diffusion models warmed up on central images can learn to generate images of different categories more easily. On {\tt CelebA} and {\tt Camelyon}, which contain colorful images, our central images are still projected into 2 clusters, which explains why the synthetic images generated by \toolname are still more useful for training classifiers. However, because of the diversity of colorful images, many sensitive images are not covered by the clusters of our central images. Therefore, diffusion models still need many training iterations to learn those uncovered sensitive images than on two gray image datasets.

\begin{table}[H]
\normalsize
\setlength{\tabcolsep}{3pt}
    \centering
    \renewcommand\arraystretch{1}
    \begin{tabular}{p{0.97\columnwidth}}
    \Xhline{1.0pt}
         \rowcolor{gray0} \noindent \textbf{Answers to RQ2}: Querying central images for warm-up training significantly promotes the DP-SGD training of diffusion models. Public images are not always useful for warm-up training, especially when a large domain shift exists between the public and sensitive datasets.\\
    \Xhline{1.0pt}
    \end{tabular}
\end{table}

\subsection{RQ3: Hyper-parameters Analysis}

This experiment investigates how the
number of queried central images (e.g., $N_c$ in Section~\ref{subsec:stage1}) for warm-up impacts the performance of \toolnameMod and \toolnameMean. We explore the numbers of central images of 0, 50, 500, 750, 1000, and 1250 for all investigated image datasets with $\varepsilon=1$.

\begin{figure}[!t]
    \centering
    \setlength{\abovecaptionskip}{0pt}
    \includegraphics[width=0.99\linewidth]{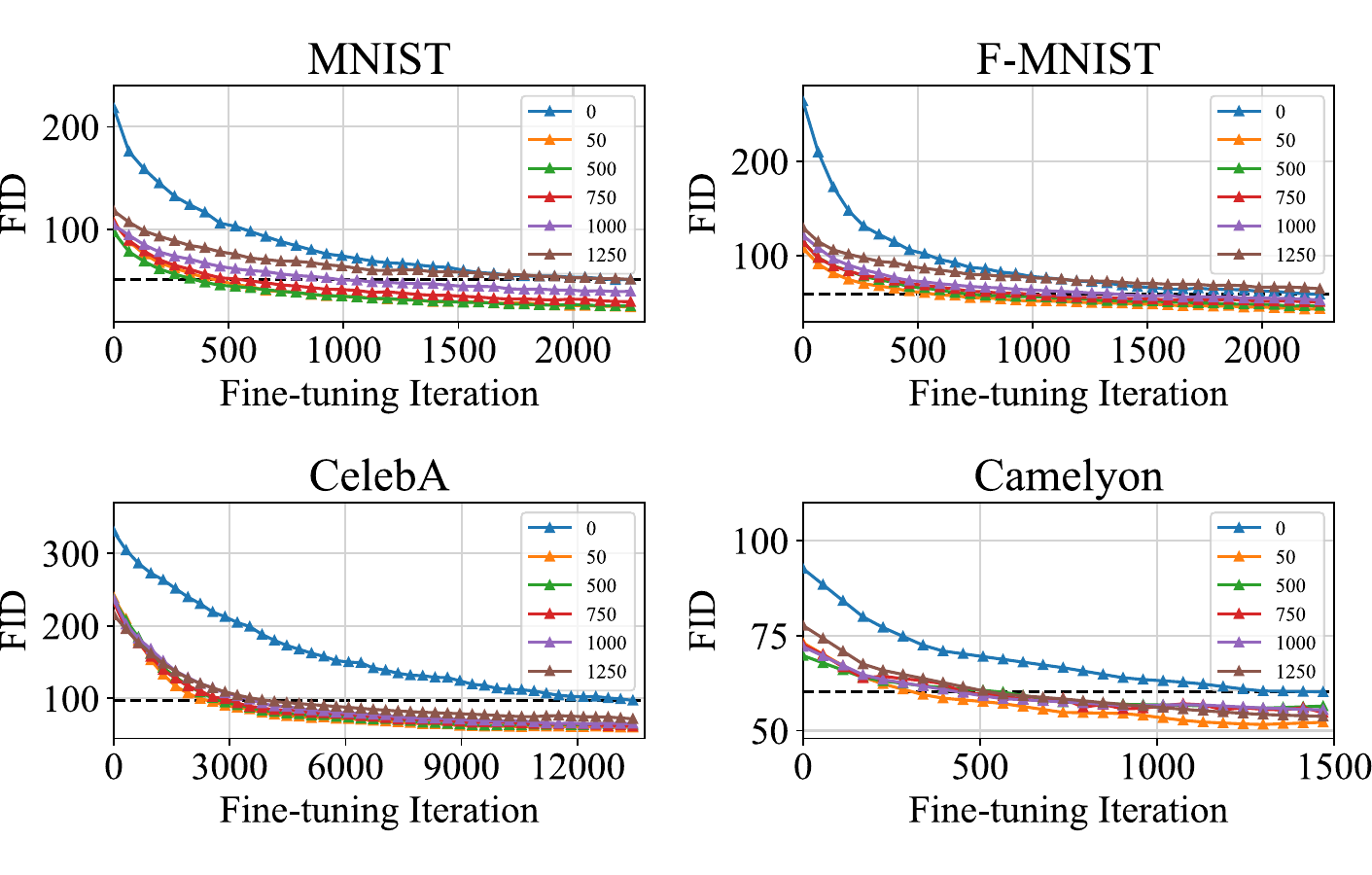}
    \caption{Convergence performance of \toolname with $\varepsilon=1$, when querying different number of mode images. The black dashed line means the FID at the end of training of diffusion models without using central images for warm-up training.}
    \label{fig:fidCurve}
\end{figure}

Figure~\ref{fig:rq3} shows that querying a small number of central images for warm-up training is a more appropriate choice for all four investigated image datasets. Specifically, on {\tt MNIST} and {\tt F-MNIST}, the best number of central images for warming-up is just 50 for both \toolnameMod and \toolnameMean. When the number of central images increases, the FID and Acc get worse. This is because, given a limited privacy budget, if we allocate too much privacy budget to querying central images, the budget allocated to private fine-tuning gets less. Therefore, we need to inject larger noise into the gradient during fine-tuning, which slows down the convergence of diffusion models. We conduct a quantitative analysis of the effect of querying central images on the noise scale of DP-SGD in Section~\ref{subsec:PrivacyCostCenter}.

On {\tt CelebA} and {\tt Camelyon}, the best number of central images for warming up increases. On {\tt CelebA}, \toolnameMod achieves the best Acc when querying 750 central images, while \toolnameMean obtains the best Acc when querying 500 images. On {\tt Camelyon}, \toolnameMod achieves the best Acc when querying 1000 central images, while \toolnameMean obtains the best Acc when querying 500 images. We consider this because when the number of sensitive images is large, or the distribution of images is complex, a small number of central images can only capture a small part of simple features, and diffusion models warmed up on these central images can not quickly learn the distribution of the sensitive data, especially those data point which are far from the cluster of central images. However, when the number of queried central images continues to increase, the performance of both \toolnameMod and \toolnameMean still drops. The best number of querying images for synthesizers warm-up indicates that \toolname only needs to warm up on a very small central image dataset, which can save computational resources compared to using millions of public images for pre-training~\cite{dpdm-sota,privimage}. We present the examples in Appendix~\ref{apsubsec:MoreSynthetic}.

Figure~\ref{fig:fidCurve} presents the FID\footnote{The FID during the private fine-tuning is calculated from 5,000 synthetic images for training efficiency and could be a little higher than our reported FID, which is calculated from 60,000 synthetic images.} of \toolname during the fine-tuning of diffusion models with $\varepsilon=1$, when querying different numbers of mode images. 
Across all datasets examined, diffusion models pre-trained on central images exhibit slower FID improvement during the initial stages of fine-tuning, suggesting that our warm-up has already learned the initial features of sensitive images before fine-tuning. On average, our \toolname achieves the same FID while only using 18\% of the fine-tuning iterations of diffusion models, which are not warmed up.

\begin{table}[H]
\normalsize
\setlength{\tabcolsep}{3pt}
    \centering
    \renewcommand\arraystretch{1}
    \begin{tabular}{p{0.97\columnwidth}}
    \Xhline{1.0pt}
         \rowcolor{gray0} \noindent \textbf{Answers to RQ3}: The optimal number of central images is significantly smaller than the number of sensitive images and may grow as the quantity and complexity of the sensitive image data distribution increase.  \\
    \Xhline{1.0pt}
    \end{tabular}
\end{table}



\section{Discussion}
\label{sec:discussion}

This section discusses (1) how \toolname performs when using public images, (2) the reason why mean is better than mode, (3) the privacy cost of central images, (4) the time cost of \toolname, and (5) the inherent limitations of \toolname.

\subsection{Combining Central Image with Public Image}
\label{subsec:dis1}

In Section~\ref{subsec:rq2}, we find that, on {\tt MNIST} and {\tt F-MNIST}, using central images for warm-up is better than using public images (e.g. {\tt ImageNet}), while on {\tt CelebA} and {\tt Camelyon}, the public images are more useful for warm-up training. This section investigates whether combining public datasets pre-training and central images warm-up can benefit synthesis performance and also compares our \toolname with existing SOTA methods using public images. We introduce six baselines as follows:

\begin{table}[!t]
\renewcommand{\arraystretch}{1.1}
\setlength{\tabcolsep}{5.5pt}
\small
    \centering
    \caption{FID and Acc of \toolnameMean, \toolnameMod and six baselines which use public images on four investigated image datasets with $\varepsilon=1$. The best performance in each column is highlighted using a bold font.}
    \label{tab:dis1}
    \resizebox{0.49\textwidth}{!}{
    \begin{tabular}{l|rc|rc|rc|rc}
    \toprule
    \multirow{2}{*}{Method} & \multicolumn{2}{c|}{{\tt MNIST}} & \multicolumn{2}{c|}{{\tt F-MNIST}}
    & \multicolumn{2}{c|}{{\tt CelebA}} & \multicolumn{2}{c}{{\tt Camelyon}}\\
    \cline{2-9}
     & \centering FID & Acc & FID & Acc & FID & Acc & FID & Acc\\
    \hline
    P-DPDM & 11.4 & 87.5 & 33.2 & 75.4 & 34.4 & 80.6 & 42.4 & 81.8\\
    PrivImage & 10.9 & 93.1 & 26.9  & 79.2  & 26.8 & 80.6 & \textbf{39.1} & 82.0\\
    PE & 50.5 & 33.7 & 32.1 & 51.3 & \textbf{22.5} & 69.8 & 69.8 & 61.2\\
    \toolnameMod & 8.2 & 96.4 & 28.8 & 80.4 & 42.8 & 83.8 & 53.0 & 82.2\\
    \toolnameMean & 8.5 & 96.5 & 25.6 & \textbf{82.1} & 41.5 & 86.5 & 52.8 & 82.8\\
    \hline
    P-DPDM$_{o}$ & 9.0 & 95.9 & 28.4 & 80.8 & 35.1 & 86.2 & 40.4 & 81.0\\
    P-DPDM$_{e}$ & 9.6 & 96.1 & 24.0 & 81.7 & 32.8 & \textbf{86.8} & 42.7 & 81.8\\
    PrivImage$_{o}$ & 9.9 & 95.5 & 25.1 & 80.1 & 27.7 & 83.3 & 43.5 & \textbf{84.4} \\
    PrivImage$_{e}$ & \textbf{8.0} & \textbf{96.6} & \textbf{23.8} & \textbf{82.1} & 27.4 & 84.7 & 46.2 & 83.1\\
    \bottomrule
\end{tabular}
}
\end{table}

\begin{itemize}[leftmargin=*]

\item \textbf{P-DPDM~\cite{dpdm-sota}:} P-DPDM trains diffusion models with a large batch size to enhance the stability and convergence speed of the model training under the noise from DP-SGD. Besides, they leverage the public dataset to pre-train diffusion models, benefiting the models from a broader knowledge base.



\item \textbf{PrivImage~\cite{privimage}:} Compared to P-DPDM, which directly uses the whole public dataset, PrivImage queries the semantics distribution of the sensitive data and selects a part of public data for pre-training.

\item \textbf{PE~\cite{api}:}
\kc{Private Evolution (PE) is an algorithm that progressively guides a foundation model to generate synthetic images similar to a private dataset without the need for fine-tuning.}

\item \textbf{P-DPDM$_{e}$:} Instead of just using {\tt ImageNet} for pre-training, P-DPDM$_{e}$ combines the queried mean images and public images from {\tt ImageNet} into one training set to pre-train diffusion models.

\item \textbf{P-DPDM$_{o}$:} Like P-DPDM$_{e}$, P-DPDM$_{o}$ combines the queried mode images and public images from {\tt ImageNet} into one training set to pre-train diffusion models.

\item \textbf{PrivImage$_{e}$:} Given pre-trained diffusion models from PrivImage, we fine-tune it on our queried mean images first and then on the sensitive images with DP-SGD.

\item \textbf{PrivImage$_{o}$:} Similar to PrivImage$_{e}$, given pre-trained diffusion models from PrivImage, we fine-tune it on our queried mode images and then fine-tune it on the sensitive images with DP-SGD.

\end{itemize}

For a fair comparison, we implement all tuning-needed baselines using the same diffusion model as \toolname. For PE, we use the pre-trained model released by Nichol et al.~\cite{improvedDM}\footnote{\url{https://github.com/openai/improved-diffusion}}, which is also trained on {\tt ImageNet} for the fair comparison. 
Table~\ref{tab:dis1} presents the FID and Acc of \toolnameMod, \toolnameMean, and above seven methods with $\varepsilon=1$. It is interesting that, on {\tt MNIST} and {\tt F-MNIST}, both \toolnameMod and \toolnameMean surpass P-DPDM and PrivImage a lot, which once again validates the superiority of central images when the sensitive images are not similar to the public images. \kc{PE only demonstrates effectiveness on the FID for two RGB image datasets due to the same limitation: when the sensitive dataset diverges too significantly from the foundation model's training data, PE struggles to generate useful synthetic images without fine-tuning~\cite{gong2025dpimagebench}}. However, it seems that using both central and public images for pre-training diffusion models is not always better than using only central images.

For the two variants of P-DPDM, on {\tt MNIST}, both P-DPDM$_{o}$ and P-DPDM$_{e}$ obtain better FID and Acc than \toolnameMod and \toolnameMean, respectively. On {\tt F-MNIST} and {\tt Camelyon}, although using public images enables \toolnameMod and \toolnameMean to obtain lower FID, their Accs decrease a little. However, on {\tt CelebA}, both FID and Acc of \toolname become better. Especially, P-DPDM$_{o}$ obtains 18.0\% lower FID and 2.4\% higher Acc. Therefore, directly mixing the central images and public images to pre-train the diffusion model could not be the best way. 

For the two variants of PrivImage, they seem to obtain better performance. PrivImage$_{e}$ obtains the lowest FID and highest Acc among all these eight methods on both {\tt MNIST} and {\tt F-MNIST}, while PrivImage$_{o}$ obtains the highest Acc on {\tt Camelyon}. Besides, both variants obtain very competitive FIDs compared to PrivImage. We consider the reason why the variants of PrivImage are better than that of P-DPDM is that the number of central images is small compared to that of public images; the benefits of central images are easily overshadowed by public images. PrivImage selects 1\% of public datasets for pre-training, which greatly reduces the size of pre-training datasets. Therefore, PrivImage$_{e}$ can better utilize the benefit of central images. However, PrivImage$_{e}$ still achieves suboptimal performance on {\tt CelebA} and {\tt Camelyon}. This may stem from its sequential pre-training on public images followed by central images. In this process, public datasets (e.g., {\tt ImageNet}) provide a broad but distant feature base, while central images shift the model toward their simpler, privacy-preserving distribution. Since the later phase dominates, the stronger influence of central images may prioritize coarse characteristics over fine details, which are more critical for {\tt CelebA} (faces) and {\tt Camelyon} (human tissues). Considering the potential of these variants, especially PrivImage$_{e}$, we believe that how to leverage two types of images together for pre-training can be a hopeful future work.



\begin{figure*}[!t]
\centering
    \setlength{\abovecaptionskip}{0pt}
    \includegraphics[width=1.0\linewidth]{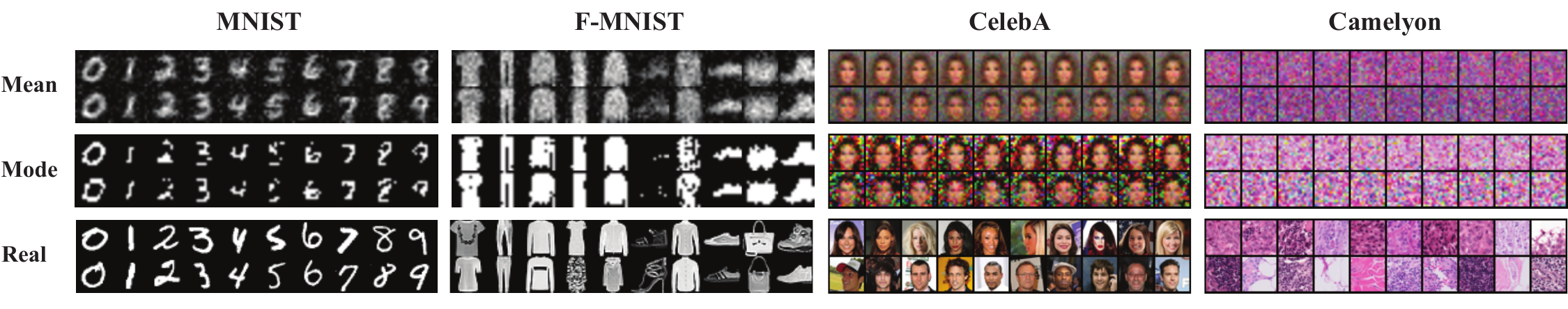}
    \caption{Examples of mean and mode images from queried from four investigated image datasets, {\tt MNIST}, {\tt F-MNIST}, {\tt CelebA} and {\tt Camelyon}. The last row of images is real image samples from each image dataset.}
    \label{fig:centerShow}
\end{figure*}

\begin{table}[!t]
\renewcommand{\arraystretch}{1.1}
\setlength{\tabcolsep}{4.2pt}
\small
    \centering
    \caption{Three metrics, FID-p, Loss-p, and FID-f (as introduced in Section~\ref{subsec:MeanVSMode}), of the diffusion model using different types of central images for warm-up training on four investigated image datasets with $\varepsilon=1$.}
    \label{tab:dis2}
    \begin{tabular}{l|c|c|c|c|c}
    \toprule
    \multicolumn{2}{c|}{Central Type} & {\tt MNIST} & {\tt F-MNIST}
    & {\tt CelebA} & {\tt Camelyon}\\
    \hline
    \multirow{3}{*}{Mode} & FID-p & 123 & 141 & 346 & 215\\
     & Loss-p & 0.19 & 0.37 & 0.32  & 0.27\\
      & FID-f & 8.2 & 28.8 & 42.8 & 53.0\\
    \hline
    \multirow{3}{*}{Mean} & FID-p & 281 & 230 & 226 & 203\\
     & Loss-p & 0.24 & 0.35 & 0.30  & 0.24\\
      & FID-f & 8.5 & 25.6 & 41.5 & 52.8\\
    \bottomrule
\end{tabular}
\end{table}

\subsection{Why is Mean Better than Mode?}
\label{subsec:MeanVSMode}

In Section~\ref{subsec:rq1}, we observe that our \toolnameMean achieves better performance than \toolnameMod on most investigated image datasets. To further explore the reason, we calculate three different metrics as follows:

\begin{itemize}[leftmargin=*]

\item \textbf{FID-p:} The FID of synthetic images generated by \toolname without fine-tuning on sensitive images and only warmed up.

\item \textbf{Loss-p:} The loss value of Equation~\ref{eq:L_DM} on the sensitive datasets from the diffusion model, which has only been warmed up like above.

\item \textbf{FID-f:} The FID of synthetic images generated by our \toolname with the two-stage training.

\end{itemize}

Table~\ref{tab:dis2} presents the three metrics on four investigated image datasets with $\varepsilon=1$. For Loss-p, on all four investigated datasets, diffusion models that are warmed up on the mean images obtain a lower loss value of the objective function than on the mode images. Thus, they learn the distribution of sensitive data from a better starting point and obtain a lower FID after private fine-tuning. For FID-p, this phenomenon seems to be consistent except for the FID-p on {\tt F-MNIST}. Specifically, the diffusion model using mode images for warm-up obtains an extremely lower FID than using mean images. However, its FID after fine-tuning gets higher than the synthesizers using mean images for warm-up. 


\begin{table}[!t]
\renewcommand{\arraystretch}{1.1}
\setlength{\tabcolsep}{5.5pt}
\small
    \centering
    \caption{The Acc of synthetic images from diffusion models warmed up with two types of central images and the noise scale $\sigma_f$ of DP-SGD on four investigated image datasets with $\varepsilon=1$, when querying different numbers of central images. The best Acc in each column is highlighted using the bold font.}
    \label{tab:dis3}
    \resizebox{0.48\textwidth}{!}{
    \begin{tabular}{lr|cc|cc|cc|cc}
    \toprule
    \multicolumn{2}{c|}{\multirow{2}{*}{Central Type}} & \multicolumn{2}{c|}{{\tt MNIST}} & \multicolumn{2}{c|}{{\tt F-MNIST}}
    & \multicolumn{2}{c|}{{\tt CelebA}} & \multicolumn{2}{c}{{\tt Camelyon}}\\
    \cline{3-10}
     & & \centering Acc & $\sigma_f$ & Acc & $\sigma_f$ & Acc & $\sigma_f$ & Acc & $\sigma_f$\\
    \hline
    \multirow{6}{*}{Mode}
    & 0 & 94.1 & 12.8 & 77.7 & 12.8 & 81.9 & 5.9 & 81.1 & 1.85\\
    &50 & \textbf{96.4} & 13.2 & \textbf{80.4}  & 13.2 & 83.7 & 6.0 & 81.4 & 1.86\\
    &500 & 96.1 & 16.1 & 78.8 & 16.1 & 83.8 & 6.8 & 82.2 & 1.91\\
    &750 & 96.2 & 19.1 & 78.5 & 19.1 & \textbf{84.2} & 7.3 & 82.0 & 1.95\\
    &1000 & 95.1 & 24.4 & 76.9 & 24.4 & 82.8 & 8.1 & \textbf{82.4} & 1.99\\
    &1250 & 93.3 & 40.0 & 75.2 & 40.0 & 81.7 & 9.2 & 81.1 & 2.05\\
    \cline{2-10}
    & Best & \multicolumn{2}{c|}{{50}} & \multicolumn{2}{c|}{{50}} & \multicolumn{2}{c|}{{750}} & \multicolumn{2}{c}{{1000}}\\
    \hline
    \multirow{6}{*}{Mean}
    & 0 & 94.1 & 12.8 & 77.7 & 12.8 & 81.9 & 5.9 & 81.1 & 1.85\\
    &50 & \textbf{96.5} & 13.2 & \textbf{82.1} & 13.2 & 84.0 & 6.0 & 82.2 & 1.86\\
    &500 & 96.1 & 16.1 & 80.7 & 16.1 & \textbf{86.5} & 6.8 & \textbf{82.8} & 1.91\\
    &750 & 96.2 & 19.1 & 79.5 & 19.1 & 85.2 & 7.3 & 81.6 & 1.95\\
    &1000 & 95.1 & 24.4 & 78.5 & 24.4 & 81.4 & 8.1 & 81.8 & 1.99\\
    &1250 & 93.3 & 40.0 & 77.4 & 40.0 & 81.2 & 9.2 & 81.2 & 2.05\\
    \cline{2-10}
     & Best & \multicolumn{2}{c|}{{50}} & \multicolumn{2}{c|}{{50}} & \multicolumn{2}{c|}{{500}} & \multicolumn{2}{c}{{500}}\\
    \bottomrule
\end{tabular}
}
\end{table}

Figure~\ref{fig:centerShow} presents the examples of two types of central images. On {\tt MNIST} and {\tt F-MNIST}, both mean and mode images depict the general shapes of different categories. On {\tt CelebA}, which contains colorful images of human faces, two types of central images depict the general outline of human faces. On {\tt Camelyon} composed of images of human tissue, although these central images do not contain any morphological features of the tissue cells, they have captured the overall color of these tissue images, pink. Our central images have captured some low-level features of the sensitive images to a greater or lesser extent, so the diffusion models pre-trained on these central images can be fine-tuned more effectively on the sensitive images. These results indicate that our central images can capture some simple and useful features for warm-up training. However, compared to mode images, mean images seem to more stably capture these useful features from sensitive images, such as the sixth column of mode images queried from {\tt F-MNIST}.

Besides, the query results of mode images are more easily affected by the injected Gaussian noise on color images. For example, the background of mode images queried from {\tt CelebA} is much noisier than that of mean images. However, since mode images are obtained through calculating the histogram of images, theoretically, querying mode data can be extended more naturally to discrete data, such as text, while querying mean data can not.


\subsection{Privacy Cost of Central Images}
\label{subsec:PrivacyCostCenter}

In \toolname, two processes consume the privacy budget: (1) querying the central images for warm-up training and (2) fine-tuning the warmed-up diffusion model using DP-SGD. This section explores how the privacy cost of querying central images impacts DP-SGD. As introduced in the Privacy Cost of Section~\ref{subsec:stage1}, we fix the privacy-related hyper-parameters except for the noise scale $\sigma_f$ of DP-SGD, and try different $\sigma_f$ to meet the required DP budget. Therefore, the number of queried central images has an impact on $\sigma_f$, which affects the private fine-tuning. 

Table~\ref{tab:dis3} presents the Acc of diffusion models using mean or mode images for a warm-up and the according  $\sigma_f$ used in DP-SGD at different numbers of queried central images. As the number of central images increases, the $\sigma_f$ also increases for less privacy budget allocated to DP-SGD, and then the gradient gets noisier. Despite more noisy gradients, diffusion models warm up on central images and still generate synthetic images of much more utility. 
The benefits brought by using central images for warm-up mitigate the performance degradation caused by DP-SGD even under more noisy gradients. For example, on {\tt CelebA}, although the noise scale of DP-SGD increases 23.7\%, the Acc of synthetic images generated by \toolname using queried mode images still increases 2.3\%. 

\begin{table}[!t]
    \centering
    \caption{Running time of \toolname.  Since \toolnameMean and \toolnameMod differ only in the query of different central images, their two stages cost the same time.}
    \begin{tabular}{l|c|c|c}
    \toprule
      Dataset & Module & \toolnameMod & \toolnameMean\\
    \hline
    \multirow{2}{*}{\shortstack{{\tt MNIST} \&\\{\tt F-MNIST}}} 
     & Stage-\uppercase\expandafter{\romannumeral1}  & 0.1h & 0.1h\\ 
     & Stage-\uppercase\expandafter{\romannumeral2}  & 12.2h & 12.2h\\
    \hline
    \multirow{2}{*}{{\tt CelebA}} 
     & Stage-\uppercase\expandafter{\romannumeral1} & 0.3h & 0.3h\\ 
     & Stage-\uppercase\expandafter{\romannumeral2} & 54.5h & 54.5h\\
    \hline
    \multirow{2}{*}{{\tt Camelyon}} 
     & Stage-\uppercase\expandafter{\romannumeral1} & 0.3h & 0.3h\\ 
     & Stage-\uppercase\expandafter{\romannumeral2} & 12.1h & 12.1h\\
    \bottomrule
\end{tabular}
\label{tab:computationalResource}
\end{table}

\subsection{Time Cost of \toolname}
\label{subsec:TimeCost}

\toolname is composed of two stages, including warm-up training and private fine-tuning. Table~\ref{tab:computationalResource} presents the time cost of \toolnameMod and \toolnameMean on these stages. Since the hyper-parameters on {\tt MNIST} and {\tt F-MNIST} are the same, the time consumption of \toolname is the same. Compared to directly fine-tuning the model using DP-SGD, the warm-up training in the Stage-\uppercase\expandafter{\romannumeral1} only introduces an average of 1.1\% additional training-time cost, while bringing a 33.1\% and 2.1\% increase in fidelity and utility metrics.



\subsection{Limitations}
\label{subsec:limitation}

The warm-up training of \toolname relies on querying central images from the sensitive dataset. The central images include two types, mean images and mode images, which need to be calculated using the mean and histogram of a randomly sampled image subset, respectively. However, if the data dimensions of sensitive images (e.g., resolution) are not all the same, we need to consider how to define the mean and mode. Besides, when the variance of the sampled subset is large, the mean and mode image could be very noisy and can hardly capture some useful features, making it ineffective for warm-up training. \kc{One potential solution is to perform DP clustering~\cite{DPClustering} on the sampled images and use the weighted centroid of the largest cluster as the central image.  
Future works should consider how to combine central images with public images for more effective pre-training.}

\section{Related Work}
\label{sec:rw}





This section discusses two main types of DP image data synthesis works, distinguishing between those that use public datasets for pre-training and those that do not. In the field of DP image synthesis, a public image dataset refers to a dataset that does not contain any data point existing in our sensitive dataset, and we can use these public images without any privacy concerns.

\noindent \textbf{Training without Pre-training.}
Based on the theoretical foundations, a private kernel means embedding estimation approach for database release~\cite{KME}, Harder et al. showed DP-MERF~\cite{dp-merf}, which uses the random Fourier features to represent each sensitive image and takes the Maximum Mean Discrepancy (MMD)~\cite{MMD} as the distribution distance between the synthetic and real datasets. Seng et al.~\cite{pearl} suggested substituting the MMD with the characteristic function distance, which uses Fourier transformation to obtain the feature vectors of sensitive images, and proposed PEARL to improve generalization capability.
Jiang et al.~\cite{dp-kernel} applied functional RDP to functions in the reproducing kernel Hilbert space to propose DP-Kernel, which achieves SOTA results on the utility of synthetic images. 
Yang et al.~\cite{dp-ntk} considered using the features of empirical neural tangent kernels to replace the random Fourier features and achieved better synthesis performance. 


The aforementioned methods exhibit poor synthetic performance on complex datasets, e.g., {\tt CelebA~\cite{celeba}}. With the rapid development of deep generative models for addressing complex image generation,  various works sanitize the training process (e.g., gradients calculation) of popular generative models, like GANs~\cite{dpsgd-gan1,dpsgd-gan2,dpsgd-gan3,dpsgd-gan4,gs-wgan,datalens,g-pate} and diffusion models~\cite{dpdm,dpldm,dpdm-sota,privimage}. Based on the Private Aggregation of Teacher Ensembles (PATE) framework, Long et al.~\cite{g-pate} proposed G-PATE, which modified the standard training process of GANs via PATE. 
Wang et al. ~\cite{datalens} proposed Datalens, which compressed the gradients before aggregation, allowing injecting less noise for better performance. Another type of method applied DP-SGD~\cite{dpsgd} to sanitize the training process of deep generative models. Dockhorn et al.~\cite{dpdm} proposed DP Diffusion Models (DPDM). To alleviate the impact of injected noise, they proposed noise multiplicity, a powerful modification of DP-SGD tailored to the training of diffusion models, and achieved SOTA performance on standard benchmarks.

\noindent \textbf{Training with Pre-training.} Recently, various methods use a non-sensitive to pre-train the existing generative models. For example, based on DP-MERF~\cite{dp-merf}, Harder et al. introduced DP-MEPF~\cite{dp-mepf}, a method that leverages the public data to transform each sensitive image into a more useful perceptual feature vector. These feature vectors calculate their random Fourier features for MMD like DP-MERF. Based on DPDM~\cite{dpdm}, Ghalebikesabi et al.~\cite{dpdm-sota} proposed to first pre-train the diffusion model on a large public image dataset and then fine-tune it on the sensitive dataset with DP-SGD. Instead of directly fine-tuning the whole diffusion model via DP-SGD, Lyu et al.~\cite{dpldm} found that fine-tuning only a small part of the parameters was more effective, especially the attention modules in the neural networks. To improve training effectiveness, Li et al.~\cite{privimage} proposed PrivImage, which queried the semantic distribution of the sensitive data to select a minimal amount of public data for pre-training. PrivImage achieved new SOTA results on common image synthesis benchmarks. 
Despite this, fine-tuning using DP-SGD still consumes a huge amount of GPU memory for its need for the sample-wise gradient.
\kc{Lin et al.~\cite{api} proposed a fine-tuning-free method, PE. PE progressively guided a diffusion model, which has been pre-trained on the public images, to generate synthetic images similar to the sensitive ones in feature space.}


\section{Conclusion}
\label{sec:conclusion}

This paper proposes a two-stage DP image synthesis framework, \toolname. Compared to directly training diffusion models on the sensitive data using DP-SGD, \toolname leverages a two-stage training, where diffusion models can learn from easy to hard. In \toolname, diffusion models can learn the distribution of sensitive data much better than existing one-stage training methods.
In order to combine \toolname with using public images, we try to use both the central images and public images for warm-up training, which is not always better than using just central images. We believe that leveraging these two types of images together for warm-up could be a hopeful future work.



\bibliographystyle{ieeetr}
\bibliography{bib}

\appendices

\setcounter{section}{0}
\renewcommand\thesection{\Alph{section}}

\section{Missing Proofs}
\label{apsec:MissingProofs}

\noindent \textbf{Proof of Theorem~\ref{the:meanImage}: \textit{The query of mean image $x^\text{mean}$ has global sensitivity $\Delta_\text{mean}=C_c/B_c^*$. For any $\alpha>1$, incorporating noise $\mathcal{N}\left(0,{\sigma_c^2 \Delta_\text{mean}^2} \mathbb{I} \right)$ into the mean image $x^\text{mean}$ makes the query results satisfies $\left(\alpha, \gamma \right)$-RDP, where $\gamma  \ge {D_\alpha }\left( {\left[ {\left( {1 - q_c} \right){p_0} + q_c{p_1}\left\| {{p_0}} \right.} \right]} \right)$.}}

\vspace{2mm}
\noindent \textit{Proof.} For any two neighboring image subsets $D_{s1}=\{x_i^s\}_{i=1}^{B_c}$ and $D_{s2}=\{x_i^s\}_{i=1}^{B_c-1}$ with $\|x_i^{s,c}\|_2 \leq C_c$, we have

\begin{equation}
    \begin{split}
        \Delta_\text{mean} =& {\left\| {x_1^\text{mean} - x_2^\text{mean}} \right\|_2}\\
        =& {\left\| \frac{1}{B_c^*} \sum_{i=1}^{B_c} {x_i^{s,c}} - \frac{1}{B_c^*}  \sum_{i=1}^{B_c-1} {x_i^{s,c}} \right\|_2}\\
        =& {\left\| \frac{1}{B_c^*} {x_i^{B_c}} \right\|_2}\\
        \leq& \frac{C_c}{B_c^*}
    \end{split}
    \nonumber
\end{equation}

\begin{table*}[!t]
    \centering
    \caption{Hyper-parameters for querying central images.}
    \label{ap:centerimages}
    \begin{tabular}{l|cccc|cccc}
    \toprule
    \multirow{2}{*}{Hyper-parameter}& \multicolumn{4}{c|}{Mean Image} & \multicolumn{4}{c}{Mode Image}\\
    \Xcline{2-9}{0.5pt}
    & {\tt MNIST}& {\tt F-MNIST}& {\tt CelebA} & {\tt Camelyon}& {\tt MNIST}& {\tt F-MNIST}& {\tt CelebA} & {\tt Camelyon}\\
    \hline
    Batch size & 6K & 6K & 12K & 12K & 6K & 6K & 12K & 12K\\
    Noise scale & 5 & 5 & 10 & 10 & 5 & 5 & 10 & 10 \\
    Image norm bound & 28.0 & 28.0 & 55.5 & 55.5 & - & - & - & -\\
    Histogram dimension & - & - & - & - & 2 & 2 & 16 & 16\\
    \bottomrule
\end{tabular}
\end{table*}

\begin{table*}[!t]
    \centering
    \caption{Hyper-parameters for training diffusion models.}
    \label{ap:diffusion_details}
    \begin{tabular}{l|cccc|cccc}
    \toprule
    \multirow{2}{*}{Hyper-parameter}& \multicolumn{4}{c|}{Warm-up Training} & \multicolumn{4}{c}{Private Fine-tuning}\\
    \Xcline{2-9}{0.5pt}
    & {\tt MNIST}& {\tt F-MNIST}& {\tt CelebA} & {\tt Camelyon}& {\tt MNIST}& {\tt F-MNIST}& {\tt CelebA} & {\tt Camelyon}\\
    \hline
    Learning rate & 3×$10^{-4}$& 3×$10^{-4}$& 3×$10^{-4}$ & 3×$10^{-4}$ & 3×$10^{-4}$ & 3×$10^{-4}$& 3×$10^{-4}$ & 3×$10^{-4}$\\
    Iterations & 2K & 2K & 2.5K & 2.5K & 2.2K & 2.2K & 13.4K & 1.5K \\
    Batch size & 64 & 64 & 64 & 64 & 4096 & 4096 & 2048 & 2048\\
    Parameters & 1.5M & 1.5M & 1.5M & 1.8M & 1.5M & 1.5M & 1.5M & 1.8M\\
    \bottomrule
\end{tabular}
\vspace{-6.0pt}
\end{table*}

\noindent \textbf{Proof of Theorem~\ref{the:modePixel}: \textit{The query of frequency histogram $H_p$ has global sensitivity $\Delta_\text{p}=1$. For any $\alpha>1$, incorporating noise $\mathcal{N}\left(0,{\sigma_c^2} \mathbb{I} \right)$ into the frequency histogram $H_p$ makes the query results satisfies $\left(\alpha, \gamma \right)$-RDP, where $\gamma  \ge {D_\alpha }\left( {\left[ {\left( {1 - q_c} \right){p_0} + q_c{p_1}\left\| {{p_0}} \right.} \right]} \right)$.}}

\vspace{2mm}
\noindent \textit{Proof. We prove that querying the frequency histogram $H_p$ has global sensitivity $\Delta_\text{p}=1$. A frequency histogram $H_p \in  [0,B_c]^{1 \times bins}$ is calculated from a pixel set $D_p = \{p_i\}_{i=1}^{B_c}$, which contains $B_c$ pixels $p_i \in \mathbb{R}$ and each pixel only contributes to adding one in $H_p$. Therefore, for two frequency histograms $H_{p1}$ and $H_{p2}$, where $H_{p1}$ is obtained by adding or removing one pixel from its pixel set $D_p1$. It is obviously that the global sensitivity $\Delta_\text{p} = {\left\| {H_{p1} - H_{p2}} \right\|_2} = 1 $.}

\vspace{2mm}
\noindent \textbf{Proof of Theorem~\ref{the:modeImage}: \textit{The query of frequency histogram of all pixels $H$ has global sensitivity $\Delta_\text{mode}=\sqrt{WHC_x}$. For any $\alpha>1$, incorporating noise $\mathcal{N}\left(0,{WHC_x \sigma_c^2} \mathbb{I} \right)$ into the frequency histogram $H$ makes the query results satisfies $\left(\alpha, \gamma \right)$-RDP, where $\gamma  \ge {D_\alpha }\left( {\left[ {\left( {1 - q_c} \right){p_0} + q_c{p_1}\left\| {{p_0}} \right.} \right]} \right)$.}}

\vspace{2mm}
\noindent \textit{Proof.} We first prove querying the frequency histogram of all pixels $H \in [0,B_c]^{WHC_x \times bins}$ has global sensitivity $\Delta_\text{mode}=\sqrt{WHC_x}$. In the proof of Theorem~\ref{the:modePixel}, we have the frequency histogram sensitivity of one pixel in the image $\Delta_\text{p} = {\left\| {H_{p1} - H_{p2}} \right\|_2} = 1 $. For two frequency histograms $H_{1}$ and $H_{2}$, where $H_{1}$ is obtained by adding or removing one image from its sensitive dataset $D_{s1}$, we have

\begin{equation}
    \begin{split}
        \Delta_\text{mode}^2 =& {\left\| {H_{1} - H_{2}} \right\|_2^2}\\
        \leq& \sum_{i=1}^{WHC_x} {\left\| {H_{p1}^i - H_{p2}^i} \right\|_2^2 }\\
        \leq& \sum_{i=1}^{WHC_x} {\left\| {H_{p1} - H_{p2}} \right\|_2^2 }\\
        =& WHC_x {\left\| {H_{p1} - H_{p2}} \right\|_2^2 }\\
        =& WHC_x
    \end{split}
    \nonumber
\end{equation}

\noindent Therefore, we have $\Delta_\text{mode} \leq \sqrt{WHC_x}$.



\section{Implementation Details}
\label{apsec:id}

This section provides detailed information on the baselines, offers an in-depth explanation of \toolname, and outlines the metrics used.

\subsection{Details of Baselines}
\label{apsubsec:Baselines}

We implement all baselines using their open-source codes in our experiments as follows:

\begin{itemize}[leftmargin=*]

\item \textbf{DP-MERF~\cite{dp-merf}:} DP-MERF uses the random feature representation of kernel mean embeddings with MMD~\cite{MMD} to minimize the distribution distance between the true data and synthetic data for DP data generation. We use their open-source codes to implement DP-MERF\footnote{\url{https://github.com/ParkLabML/DP-MERF}}. We conduct experiments on {\tt Camelyon} using the same hyper-parameter setting as their setting on {\tt CelebA}.

\item \textbf{G-PATE~\cite{g-pate}:} G-PATE leverages generative adversarial nets~\cite{gan} to generate data. They train a student data generator with an ensemble of teacher discriminators and propose a novel private gradient aggregation mechanism to ensure DP. We use their open-source codes to implement G-PATE\footnote{\url{https://github.com/AI-secure/G-PATE}}. We conduct experiments on {\tt Camelyon} using the same hyper-parameter setting as their setting on {\tt CelebA}.

\item \textbf{DataLens~\cite{datalens}:} To further accelerate the convergence of data generator in G-PATE, DataLens introduces a novel dimension compression and aggregation approach, which exhibits a better trade-off on privacy and convergence rate. We use their open-source codes to implement DataLens\footnote{\url{https://github.com/AI-secure/DataLens}}. We conduct experiments on {\tt Camelyon} using the same hyper-parameter setting as their setting on {\tt CelebA}.

\item \textbf{DP-Kernel~\cite{dp-kernel}:} DP-Kernel develops the functional RDP and privatizes the loss function of data generator in a reproducing kernel Hilbert space for DP image synthesis. We use their open-source codes to implement DataLens\footnote{\url{https://github.com/dihjiang/DP-kernel}}. We conduct experiments on {\tt Camelyon} using the same hyper-parameter setting as their setting on {\tt CelebA}.


\item \textbf{DPDM~\cite{dpdm}:} DPDM trains the diffusion models on sensitive images with DP-SGD~\cite{dpsgd}. They propose noise multiplicity, a modification of DP-SGD, to alleviate the impact of injected noise to gradients. We use their open-source codes to implement DataLens\footnote{\url{https://github.com/nv-tlabs/DPDM}}. The hyper-parameters of training are the same as present in Table~\ref{ap:centerimages}.


\end{itemize}

\subsection{Details of Querying Central Image}
\label{apsubsec:CenterDetails}

Table~\ref{ap:centerimages} presents the hyper-parameters of querying central images used in our experiments. Since two colorful image datasets contain more images, which means we can obtain a smaller sample ratio given the same batch size, we use a larger batch size on {\tt CelebA} and {\tt Camelyon} than {\tt MNIST} and {\tt F-MNIST}. We set the image norm bound as the upper bound of images. Since all the sensitive images are scaled into $[0,1]^{W \times H \times C_x}$, where $W$ and $H$ are the width and height of the image respectively, and $C_x$ is the number of color channels, the image norm is always smaller than $\sqrt{W \times H \times C_x}$. The $W \times H \times C_x$ of {\tt MNIST} and {\tt F-MNIST} is $28 \times 28 \times 1$, and that of {\tt CelebA} and {\tt Camelyon} is $32 \times 32 \times 3$. For the histogram dimension of mode images, we use 2 and 16 for gray image datasets and colorful image datasets, respectively.


\subsection{Details of Augmentation}
\label{apsubsec:AugDetails}

We implement the augmentation algorithm bag as introduced in~\ref{subsubsec:pre-training} using 14 image augmentation algorithms proposed by Cubuk et al.~\cite{randomAug}, which can be accessed at the repository\footnote{\url{https://github.com/tensorflow/tpu/blob/master/models/official/efficientnet/autoaugment.py}}. For all investigated image datasets, we randomly sample 2 augmentation
algorithms from the bag to sequentially transform the input central images during the pre-training.

\subsection{Details of Model Training}
\label{apsubsec:TrainDetails}

Table~\ref{ap:diffusion_details} presents the hyper-parameters of warm-up training and private fine-tuning used in our experiments. All the hyper-parameters of fine-tuning are the same as DPDM~\cite{dpdm}, and we find their setting works well. For warm-up training, we use the same learning rate on all four investigated image datasets. Since the number of queried central images is small, we warm up diffusion models for a small number of iterations, and use a small batch size. 

\subsection{Details of Metrics}
\label{apsubsec:MetricDetails}


\begin{itemize}[leftmargin=*]

\item  \textbf{Fr\'{e}chet Inception Distance (FID):} FID has been widely used to assess the fidelity of synthetic images generated by Generative models~\cite{ddpm,ddim,biggan}. A lower FID suggests that the generated images are higher quality and more akin to the real dataset. We generate 60,000 synthetic images to calculate FID.

\item \textbf{Acc:} We assess the utility of synthetic images on the image classification task. Following DPDM, we train a CNN classifier on the synthetic images, and the Acc is tested on the sensitive test dataset. We generate 60,000 synthetic images to train the classifier.

\end{itemize}

For FID, we use the pre-trained Inception V1\footnote{\url{https://api.ngc.nvidia.com/v2/models/nvidia/research/stylegan3/versions/1/files/metrics/inception-2015-12-05.pkl}} as DPDM~\cite{dpdm} to extract the feature vectors of images. For Acc, the model architecture of the CNN classifier is taken from the repository~\footnote{\url{https://github.com/nv-tlabs/DP-Sinkhorn_code}} as DPDM.

\section{More Results}
\label{apsec:moreResults}

\subsection{Visualization Comparison}
\label{apsubsec:vc}

Examples of synthetic images from DP-Kernel~\cite{dp-kernel}, DPDM~\cite{dpdm}, \toolnameMod and \toolnameMean on {\tt MNIST}, {\tt F-MNIST}, {\tt CelebA}, and {\tt Camelyon} are present in Figure~\ref{fig:rq1_e10} with $\varepsilon=10$. With a larger privacy budget ($1 \rightarrow 10$), the fidelity of synthetic images from DPDM is competitive with ours. However, on more complex datasets, {\tt CelebA} and {\tt Camelyon}, our \toolname still surpasses DPDM. For example, DPDM sometimes fails to generate complete facial images (the second row), and on {\tt Camelyon}, DPDM generates some very similar images, reducing the diversity of the synthetic dataset. In spite of an increase in the privacy budget, the images generated by DP-Kernel are still blurry across all four datasets.





\begin{table}[!t]
\renewcommand{\arraystretch}{1.1}
\setlength{\tabcolsep}{5.5pt}
\footnotesize
    \centering
    \caption{FID, Acc and $\sigma_f$ of \toolnameMean on {\tt MNIST} and {\tt F-MNIST} under $\varepsilon=1$ with two different privacy accountant, RDP and PRV.}
    \label{tab:prv}
    \resizebox{0.4\textwidth}{!}{
    \begin{tabular}{l|ccc|ccc}
    \toprule
    \multirow{2}{*}{Account} & \multicolumn{3}{c|}{{\tt MNIST}} & \multicolumn{3}{c}{{\tt F-MNIST}}\\
    \cline{2-7}
     & \centering FID & Acc & $\sigma_f$ & FID & Acc & $\sigma_f$\\
    \hline
    RDP & 8.0 & 96.6 & 13.2 & 23.8 & 82.1 & 13.2\\
    PRV & 7.9 & 96.3 & 12.2 & 23.6 & 82.5 & 12.2\\
    \bottomrule
\end{tabular}
}
\end{table}

\subsection{More Synthetic Images from \toolname}
\label{apsubsec:MoreSynthetic}

Figure~\ref{fig:exampleMeanDiffNumber} and~\ref{fig:exampleModeDiffNumber} present the synthetic images from \toolnameMean and \toolnameMod, when querying different numbers of central images for pre-training. On {\tt MNIST} and {\tt F-MNIST}, two variants of \toolname, achieves suboptimal synthesis performance when the number of queries is too small or too large. On two colorful image datasets, {\tt CelebA} and {\tt Camelyon}, the diversity of the synthetic images is poor when no central images are used for pre-training. For example, some synthetic {\tt Camelyon} images are very similar to each other. When the number of sensitive images is large, querying more central images could be better.

\subsection{More Advanced Privacy Accounting}
\label{apsubsec:prv}

\kc{In this paper, we use the RDP to track the privacy cost of \toolname for fair comparison with existing methods. We also explore using a more advanced privacy accountant, Privacy loss Random Variables (PRV)~\cite{prv}. As shown in Table~\ref{tab:prv}, the noise scale $\sigma_f$ of \toolnameMean can be reduced by 7.8\% with PRV. With less noisy gradients, \toolnameMean performs better, but this improvement is slight.}

\begin{figure*}[!t]
    \centering
    \setlength{\abovecaptionskip}{5pt}
    \includegraphics[width=1.0\linewidth]{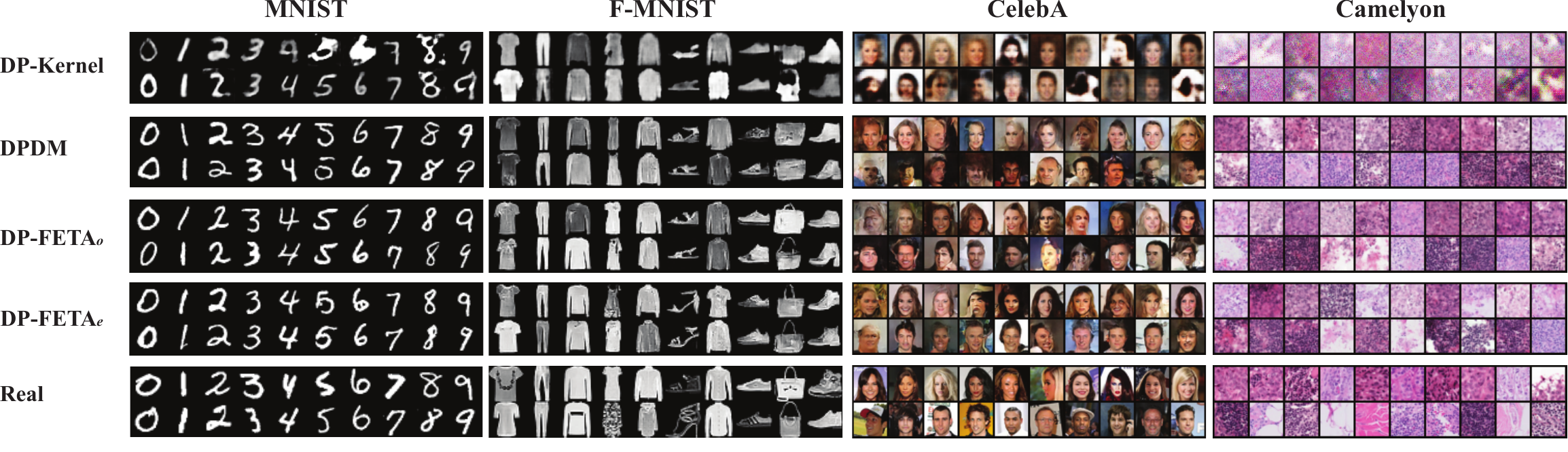}
    \caption{Examples of synthetic images from four different methods, DP-Kernel~\cite{dp-kernel}, DPDM~\cite{dpdm} and our \toolnameMod and \toolnameMean, on four investigated image datasets, {\tt MNIST}, {\tt F-MNIST}, {\tt CelebA} and {\tt Camelyon}, with $\varepsilon=10$. The last row of images are real image samples from each image dataset.}
    \label{fig:rq1_e10}
    \vspace{10.0pt}
\end{figure*}

\begin{figure*}[!t]
    \centering
    \setlength{\abovecaptionskip}{5pt}
    \includegraphics[width=1.0\linewidth]{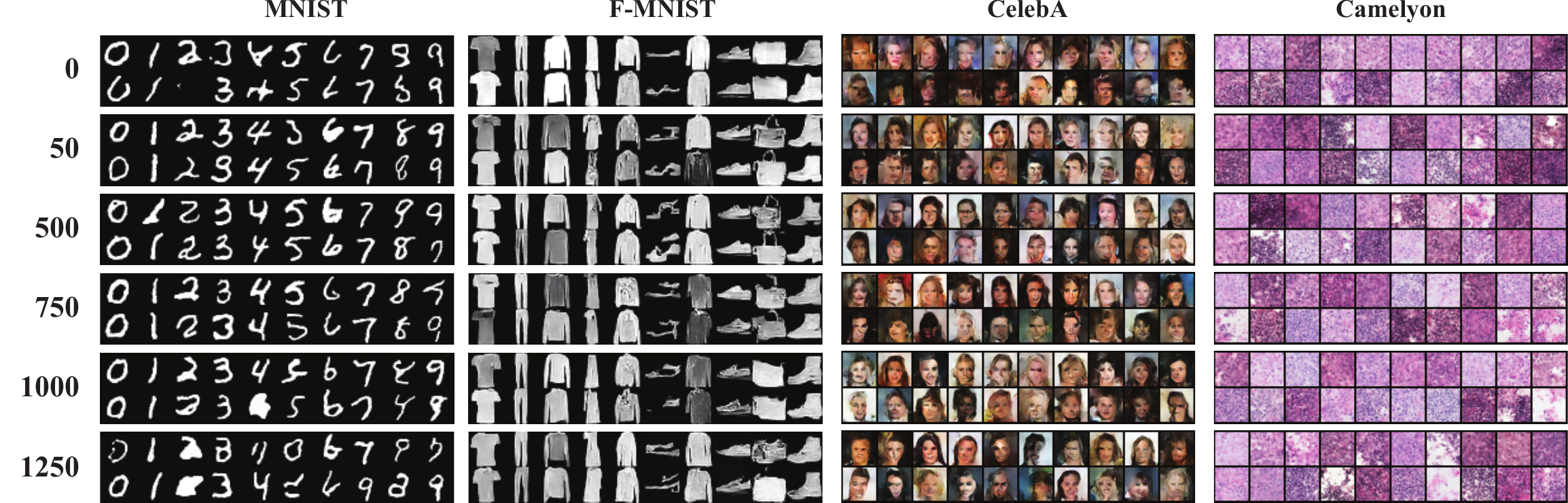}
    \caption{Examples of synthetic images from \toolnameMean on four investigated image datasets, {\tt MNIST}, {\tt F-MNIST}, {\tt CelebA} and {\tt Camelyon}, with $\varepsilon=1$, when querying different numbers of mean images.}
    \label{fig:exampleMeanDiffNumber}
    \vspace{10.0pt}
\end{figure*}

\begin{figure*}[!t]
    \centering
    \setlength{\abovecaptionskip}{5pt}
    \includegraphics[width=1.0\linewidth]{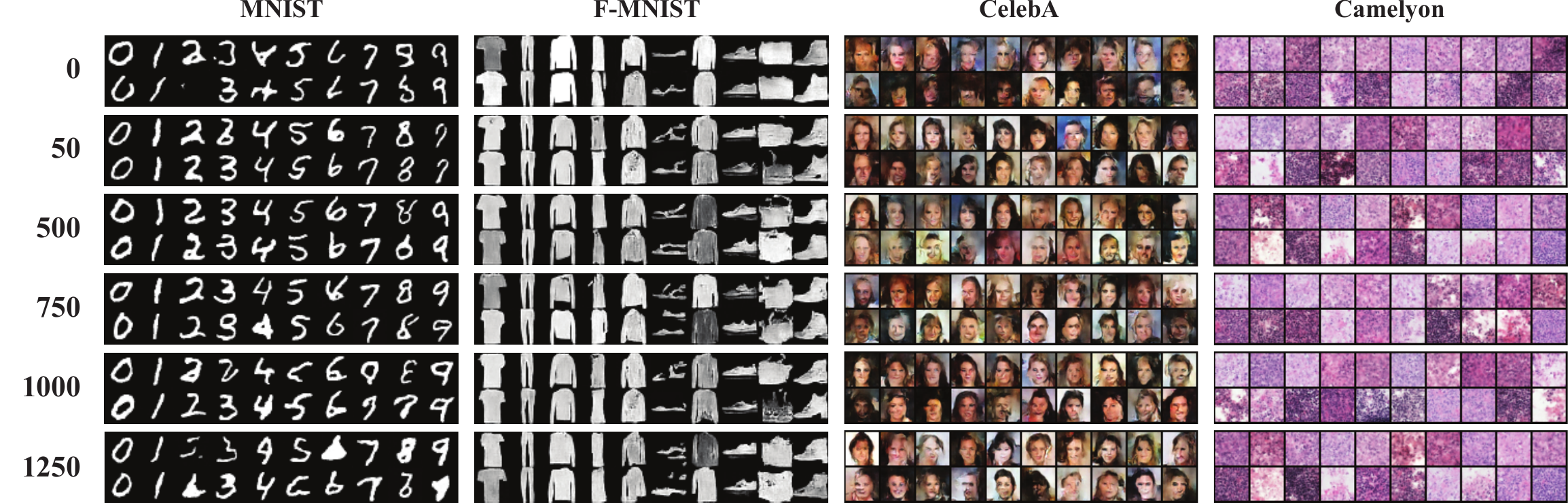}
    \caption{Examples of synthetic images from \toolnameMod on four investigated image datasets, {\tt MNIST}, {\tt F-MNIST}, {\tt CelebA} and {\tt Camelyon}, with $\varepsilon=1$, when querying different numbers of mode images.}
    \label{fig:exampleModeDiffNumber}
    \vspace{10.0pt}
\end{figure*}

\clearpage


\section{Meta-Review}

The following meta-review was prepared by the program committee for the 2025
IEEE Symposium on Security and Privacy (S\&P) as part of the review process as
detailed in the call for papers.

\subsection{Summary}
This paper proposes a new method for training differentially-private machine learning models. The core idea is to first train models on aggregate DP features of the dataset. Then, they fine-tune using SGD. The approach demonstrates a better privacy-utility tradeoff than prior methods.

\subsection{Scientific Contributions}
\begin{itemize}
\item Establishes a new research direction
\item Provides a Valuable Step Forward in an Established Field
\end{itemize}

\subsection{Reasons for Acceptance}
\begin{enumerate}
\item This approach is interesting and novel. To our knowledge, it has not be tried before.
\item The empirical results appear to be rather promising.
\end{enumerate}




\end{document}